\documentclass[a4paper,11pt]{article}

\usepackage[utf8]{inputenc}
\usepackage[english]{babel}
\usepackage[a4paper]{geometry}
\usepackage{amsfonts}
\usepackage{amsmath}
\numberwithin{equation}{section}
\usepackage{bm}
\usepackage{amsthm}
\usepackage{amssymb}
\usepackage{mathrsfs}
\usepackage{bbm}
\usepackage{enumitem}
\usepackage{empheq}
\usepackage{graphicx}
\usepackage{xcolor}
\usepackage{placeins}
\usepackage{multirow}
\usepackage{colortbl}
\usepackage{booktabs}
\usepackage{natbib}
\usepackage{doi}

\usepackage{hyperref}
\definecolor{Green4}{rgb}{0,.2,0}
\hypersetup{colorlinks, linkcolor=red, citecolor=blue, urlcolor=Green4}

\usepackage{subcaption}
\usepackage{algorithm}
\usepackage{algorithmic}

\DeclareMathOperator{\Tr}{Tr}

\DeclareMathOperator{\Diag}{Diag}

\DeclareMathOperator*{\argmin}{arg\,min}
\DeclareMathOperator*{\argmax}{arg\,max}

\theoremstyle{plain}
\newtheorem{thm}{Theorem}
\newtheorem{prop}{Proposition}
\newtheorem{lemma}{Lemma}

\theoremstyle{definition}

\title{Multi-fidelity Gaussian process regression for noisy outputs and non-nested experimental designs: decoupled versus joint likelihood optimization}
\author{Nils Baillie$^{1,2,}$\footnote{Corresponding author: \href{mailto:nils.baillie@cea.fr}{nils.baillie@cea.fr}} , Baptiste Kerleguer$^{3,4}$, Cyril Feau$^1$, Josselin Garnier$^2$
}
\date{$^1$Universit\'e Paris-Saclay, CEA, Service d’\'Etudes Mécaniques et Thermiques, \\
91191 Gif-sur-Yvette, France \\
$^2$CMAP, CNRS, \'Ecole polytechnique, Institut Polytechnique de Paris, \\
91120 Palaiseau, France \\
$^3$CEA, DAM, DIF, F-91297, Arpajon, France \\
$^4$Universit\'e Paris-Saclay, CEA, LMCE, F-91680 Bruyères-le-Châtel, France}

\begin{document}

\maketitle

\begin{abstract}

This paper investigates a recursive formulation of auto-regressive multi-fidelity Gaussian process regression in the challenging setting of noisy and non-nested high- and low-fidelity data. We propose a decoupled optimization strategy based on the expectation-maximization algorithm, which exploits the structure of the recursive model. In particular, we derive closed-form update formulas when the scaling factor is modeled as a parametric linear predictor. This approach is compared with the fully coupled likelihood maximization of the classical, non-recursive formulation. A series of benchmark experiments, covering applications of increasing complexity, highlights the performance of both approaches. The results demonstrate that the proposed recursive strategy substantially reduces training time when large low-fidelity datasets are available, while maintaining competitive predictive accuracy and uncertainty estimation.

\end{abstract}


\section{Introduction}

Gaussian process (GP) regression (\cite{rasmussen2005, gramacy2020surrogates}), also referred to as kriging, is a widely used tool in surrogate modeling because of its ability to provide a flexible predictor with a quantification of the prediction uncertainty in closed-form. We consider in this paper the multi-fidelity (MF) framework, where observations from two computer codes or experiments of different levels of fidelity are available. The high-fidelity (HF) data is more accurate but costly to obtain, whereas the low-fidelity (LF) data is cheaper but less accurate. A MF surrogate model exploits both low- and high-fidelity data points to approximate the target HF function. \cite{kennedy2000ohagan} proposed the auto-regressive or AR(1) MFGP model in which the HF function is the LF function multiplied by a scaling term plus a discrepancy function. This paper addresses parameter estimation for this model when the outputs are noisy at both levels and when the input sets are not nested, i.e. the HF design is not a subset of the LF design. \\

In the original, non-recursive formulation (\cite{kennedy2000ohagan}), the HFGP is conditioned on the data of all fidelity levels and a large covariance matrix must be inverted. Under noise-free outputs and nested designs, the likelihood can be factorized such that the parameters of each level can be estimated separately (\cite{kennedy2000ohagan, Forrester2007function, legratiet2013bayeshierarchicalMF}). \cite{legratiet2014rec_cokriging} proposed the recursive formulation in which the HFGP is defined in terms of the posterior LFGP. This yields a surrogate model at each level, conditioned on the data of this level and of the lower ones, and only smaller matrices have to be inverted. Maximum likelihood estimation (MLE), partially Bayesian and leave-one-out (LOO) estimation of the parameters remain tractable. A fully Bayesian treatment is given in \cite{ma2020}. Both formulations rely on noise-free outputs and nested designs, which can be restrictive in sequential design (\cite{legratiet2015sequential}) for instance. For non-nested designs or noisy LF outputs, the likelihood no longer separates into one term per level. For non-nested designs, more general co-kriging equations are given by \citet[p.~274]{legratiet_thesis} and the direct numerical maximization of the likelihood over all HF parameters yields no explicit estimators. An EM algorithm (\cite{Dempster1977EM, zertuche_thesis}) replaces this maximization by a sequence of simpler problems that increase the likelihood at each iteration. All the previously mentioned articles on MFGPs suppose that the outputs are noise-free, thereby ignoring measurement errors in experiments for example. Noisy outputs at both fidelity levels have been considered in the non-recursive formulation, for instance in \cite{Babaee2020CapeCod}, at the price of a joint optimization over all parameters. According to \cite{Babaee2020CapeCod}, this high-dimensional joint optimization is prohibitive for large datasets. This limits the applicability of the non-recursive formulation, and motivates a generalization of the recursive formulation to noisy outputs and non-nested designs.  \\

Our contributions are the following. We propose a decoupled optimization strategy for the LF and HF parameters that exploits the recursive formulation of the MFGP model in the case where homoscedastic Gaussian noise affects both fidelity levels and where the experimental designs are not nested. The LF parameters are estimated by MLE, and the HF parameters are optimized separately with the EM algorithm, which reduces the dimension of the optimization problem. We extend the results of \cite{zertuche_thesis} to the case of noisy outputs and when the scaling factor is a parametric linear predictor function, and obtain a closed-form update EM formula for the latter. We show that the recursive and non-recursive formulations give the same predictive distribution for fixed parameters in this general framework, so that their differences come from estimation alone. The two estimation strategies do not maximize the same criterion: the non-recursive formulation maximizes the joint likelihood of the LF and HF data, whereas ours is a two-stage procedure in which the LF parameters are only estimated from the LF data and fixed afterwards. We compare both formulations in terms of predictive performance and training time on several examples: an analytical 1D function (\cite{giannoukou2025MFPCE}), the 4D Park function (\cite{Cox2001Parkfunction, Xiong2013test_fcts}) and the real-world dataset of sea surface temperature measurements of \cite{Babaee2020CapeCod}. \\

We consider the bi-fidelity case but the approach extends in principle to more than two levels as long as each level depends only on the previous one. Parameters are plugged in at their estimated values. Accounting for their uncertainty is beyond the scope of this paper. More flexible models (\cite{Gori2025MF}) such as nonlinear auto-regressive GPs (\cite{perdikaris2017NARGP}), deep GPs (\cite{cutajar2018deepgpmf}) or GP-Bayesian neural networks (\cite{kerleguer2024}) exist but are more costly to train, and are not considered here.  \\

The paper is organized as follows. First, we recall noisy GP regression for a single fidelity level and introduce the notations in Section \ref{sec:GPR}. We present in Section \ref{sec:MFGP_model} the two MFGP formulations and their estimation strategies. Then, several performance metrics are given in Section \ref{sec:UQ_metrics}, and finally, in Section \ref{sec:Num_applications}, we present the numerical applications. Mathematical details are given in Appendices \ref{app:EM_details} and \ref{app:proof_thm}.

\section{Noisy Gaussian process regression}
\label{sec:GPR}

We are interested in approximating a deterministic target function $y : \mathbb{R}^D \rightarrow \mathbb{R}$ using only noisy observations
$z(\textbf{x}) = y(\textbf{x}) + \varepsilon(\textbf{x})$, where $\varepsilon$ is a Gaussian white noise with variance $\sigma^2_\varepsilon$, i.e. a zero-mean GP with covariance function
$\sigma^2_\varepsilon \mathbbm{1}(\textbf{x}=\textbf{x}')$. We have at our disposal a training set $\mathcal{D} = (X_{tr}, \textbf{z}_{tr})$, composed of a set of inputs $X_{tr} = (\textbf{x}_1,\dots,\textbf{x}_N) \in \mathbb{R}^{N \times D}$ and a set of observed outputs $\textbf{z}_{tr}= (z_1,\dots,z_N) \in \mathbb{R}^{N \times 1}$ such that for all $i  \in \{1,\dots,N\}$, $z_i = y(\textbf{x}_i) + \varepsilon(\textbf{x}_i)$. The prior model for $y$ is a GP $Y$ with a mean function $m$ and a covariance function $k$. We suppose that $Y$ and $\varepsilon$ are independent. We define $Z = Y+\varepsilon$, which is a GP since the sum of two independent GPs is a GP.  \\

In practice, the user has to choose the mean and covariance functions of the GP. In the following, we suppose that the mean $m$ is a linear predictor function: $m(\textbf{x})=\textbf{f}(\textbf{x})^\top \boldsymbol\beta$, where $\boldsymbol\beta \in \mathbb{R}^{p \times 1}$ is a column vector of parameters and $\textbf{f}(\textbf{x}) = (f_1(\textbf{x}),\dots,f_p(\textbf{x}))^\top$ is a vector containing a family of real-valued functions. The covariance is of the form: $k(\textbf{x},\textbf{x}') = \sigma^2 r(\textbf{x},\textbf{x}';\boldsymbol\theta)$ where $\sigma^2 > 0$ is the variance of $Y$ and $\boldsymbol\theta \in (0, \infty)^D$ is the length scale vector which controls the correlation between $Y(\textbf{x})$ and $Y(\textbf{x}')$ depending on $\textbf{x}$ and $\textbf{x}'$.
For multi-dimensional inputs $\textbf{x}=(x^{(1)},\dots,x^{(D)})$ the correlation function $r$ between vectors is a product of one-dimensional correlation functions: 
\[  r(\textbf{x},\textbf{x}';\boldsymbol\theta) = \prod_{d=1}^D r_d(x^{(d)},x'^{(d)}; \theta_d).  \]
We also define the parameter $\eta = \sigma^2_\varepsilon / \sigma^2$ which is the ratio of the noise variance and the kernel variance, this parametrization is notably used in \cite{Gu2018robustgasp} and simplifies the algebra associated to the parameter optimization in the recursive case (see below). Let $\textbf{x} \in \mathbb{R}^D$, we want to derive the mean and variance prediction at $\textbf{x}$ of the GP knowing the training data. We introduce the notations:
\[ 
\begin{cases}
F =  (f_j(\textbf{x}_i))_{(i,j) \in \{1,\dots, N\} \times\{1,\dots, p\}}  \in \mathbb{R}^{N \times p} \\
\textbf{Z}(X_{tr})= (Z(\textbf{x}_1),...,Z(\textbf{x}_N))^\top  \\
R(\boldsymbol\theta) = r(X_{tr},X_{tr};\boldsymbol\theta) = (r(\textbf{x}_i,\textbf{x}_j;\boldsymbol\theta))_{(i,j) \in \{1,\dots,N \}^2} \in \mathbb{R}^{N \times N}  \\
\textbf{r}(\textbf{x}; \boldsymbol\theta) = r(X_{tr},\textbf{x};\boldsymbol\theta) =(r(\textbf{x}_1,\textbf{x};\boldsymbol\theta),\dots,r(\textbf{x}_N,\textbf{x};\boldsymbol\theta))^\top  \in \mathbb{R}^{N\times 1}.
\end{cases}  
\]
The matrix $F$ is assumed to have full rank and $R(\boldsymbol\theta) + \eta I_N$ is symmetric positive definite (SPD) with $I_N$ being the $N \times N$ identity matrix. By the Gaussian conditioning theorem, 
the posterior distribution of the process $Y$ given $\textbf{Z}(X_{tr})=\textbf{z}_{tr}$ is the one of a GP $\widetilde{Y}$ with mean function $m_Y$ and covariance function $v_Y$ that have closed-form expressions: 
\[  
\begin{cases}    
m_Y(\textbf{x}) = \textbf{f}(\textbf{x})^\top \boldsymbol\beta + \textbf{r}(\textbf{x}; \boldsymbol\theta)^\top (R(\boldsymbol\theta) + \eta I_N)^{-1}(\textbf{z}_{tr} - F\boldsymbol\beta) \\
v_Y(\textbf{x},\textbf{x}') = \sigma^2 \!\left(r(\textbf{x},\textbf{x}';\boldsymbol\theta) - \textbf{r}(\textbf{x};\boldsymbol\theta)^\top (R(\boldsymbol\theta) + \eta I_N)^{-1}\textbf{r}(\textbf{x}';\boldsymbol\theta) \right)\!. 
\end{cases}  
\]
In the noise-free case, i.e. when $\sigma^2_\varepsilon=0$, we observe that for all $\textbf{x}_i$ in the input training set $X_{tr}$, we have that $m_Y(\textbf{x}_i) = z_i$ and $v_Y(\textbf{x}_i,\textbf{x}_i) = 0$, hence, $\widetilde{\mathbf{Y}}(X_{tr}) = \textbf{z}_{tr}$ almost surely. We refer to this fact as the ``interpolation property". This property is lost as soon as a noise term is present. \\

For the parameter optimization, we opt for the MLE, but other alternatives exist, such as Bayesian estimation (\cite{Gu2018robustgasp}), or LOO cross-validation estimation (\cite{BACHOC2013CV_GP}). With the GP prior assumption, we have that: $\textbf{Z}(X_{tr}) \sim \mathcal{N}\!\left(F\boldsymbol\beta, \sigma^2(R(\boldsymbol\theta)+\eta I_N)\right)$. We fix ($\boldsymbol\theta, \eta$) and maximize the likelihood with respect to $\boldsymbol\beta$ and then $\sigma^2$. We obtain the estimates:
\[ 
\begin{cases}
\hat{\boldsymbol\beta} = (F^{\top} (R(\boldsymbol\theta)+\eta I_N)^{-1}F)^{-1} F^{\top}(R(\boldsymbol\theta)+\eta I_N)^{-1}\textbf{z}_{tr} \\
\hat{\sigma}^2 = (\textbf{z}_{tr} - F\hat{\boldsymbol\beta})^{\top}(R(\boldsymbol\theta)+\eta I_N)^{-1}(\textbf{z}_{tr} - F\hat{\boldsymbol\beta}) \, / \, N.
\end{cases}
\]
Plugging these formulas into the log-likelihood yields a function of $(\boldsymbol\theta, \eta)$:
\[ 
\log \mathcal{L}(\textbf{z}_{tr}\, | \,\boldsymbol\theta, \eta ) = 
-\frac{N}{2}\log(\hat{\sigma}^2) - \frac{1}{2} \log \det (R(\boldsymbol\theta)+\eta I_N) - \frac{N}{2}(1+\log(2\pi)).
\]
In practice, we maximize the previous function numerically using the L-BFGS-B algorithm (\cite{zhu1997L_BFGS_B}) which is a quasi-Newton method that can heavily depend on the initialization. This is why a multi-start strategy is used. Moreover, we systematically use the Cholesky decomposition when computing terms involving the inverse or the determinant of $R(\boldsymbol\theta)+\eta I_N$, exploiting its SPD structure. This decomposition has a time complexity of $\mathcal{O}(N^3)$ and is the most expensive step in the computations. The L-BFGS-B algorithm uses the gradient of the log-likelihood, which takes the form (\cite{rasmussen2005}):
\[ 
\frac{\partial\log \mathcal{L}}{\partial \omega}(\textbf{z}_{tr}\, | \,\boldsymbol\theta, \eta ) = \frac{1}{2}\Tr \!\left( \!\left( \kappa \kappa^\top  - (R(\boldsymbol\theta)+\eta I_N)^{-1}\right) \!\frac{\partial}{\partial \omega}\!\left(R(\boldsymbol\theta)+\eta I_N \right) \right)\!,
\]
where $\kappa = (R(\boldsymbol\theta)+\eta I_N)^{-1}(\textbf{z}_{tr} - F\hat{\boldsymbol\beta}) / \sqrt{\hat{\sigma}^2}$ and $\omega \in \{ \theta_1,\dots, \theta_D, \eta \}$. 

\section{Multi-fidelity Gaussian process model}\label{sec:MFGP_model}

We consider now the case with two fidelity levels:
\[  
\begin{cases}
z_L(\textbf{x}) = y_L(\textbf{x}) + \varepsilon_L(\textbf{x})\\
z_H(\textbf{x}) = y_H(\textbf{x}) + \varepsilon_H(\textbf{x}),
\end{cases}
\]
where $z_L$ and $z_H$ are the observed functions tainted with noise, and $y_L$, $y_H$ are the target functions, supposed to be deterministic. We observe the LF training data $\mathcal{D}_L = (X_{tr}^L,\textbf{z}_{tr}^L)$ and HF training data $\mathcal{D}_H = (X_{tr}^H,\textbf{z}_{tr}^H)$. In particular, we have the input training sets $X_{tr}^L \in \mathbb{R}^{N_L \times D}$ and $X_{tr}^H \in \mathbb{R}^{N_H \times D}$, as well as the noisy outputs
$\textbf{z}_{tr}^L \in \mathbb{R}^{N_L \times 1}$ and $\textbf{z}_{tr}^H \in \mathbb{R}^{N_H \times 1}$.
We have that $\textbf{z}_{tr}^L = \textbf{z}_L(X_{tr}^L) = (z_L(\textbf{x}))_{\textbf{x} \in X_{tr}^L}$ and $\textbf{z}_{tr}^H = \textbf{z}_H(X_{tr}^H) =(z_H(\textbf{x}))_{\textbf{x} \in X_{tr}^H}$. We typically have $N_L > N_H$. The noise terms $\varepsilon_L$ and $\varepsilon_H$ 
are taken as independent GPs of mean zero and covariance functions:
\[ 
\begin{cases}
\text{Cov}(\varepsilon_L(\textbf{x}), \varepsilon_L(\textbf{x}'))=\sigma^2_{\varepsilon,L} \mathbbm{1}(\textbf{x}=\textbf{x}') \\
\text{Cov}(\varepsilon_H(\textbf{x}), \varepsilon_H(\textbf{x}'))=\sigma^2_{\varepsilon,H} \mathbbm{1}(\textbf{x}=\textbf{x}').
\end{cases} 
\]

\subsection{Recursive formulation and decoupled EM-based optimization}\label{sec:MF_AR1}

In the recursive formulation, the LF part of the MF surrogate model is identical to the single-fidelity GP model presented in the previous section, we use the same notations as those in Section \ref{sec:GPR}. The prior model for the LF function $y_L$ is a GP $Y_L$ with mean $m_L(\textbf{x}) = \textbf{f}_L(\textbf{x})^\top \boldsymbol\beta_L$ and covariance $k_L(\textbf{x},\textbf{x}')=\sigma^2_L r_L(\textbf{x},\textbf{x}';\boldsymbol\theta_L)$.
We let $Z_L = Y_L + \varepsilon_L$, with $Y_L$ and $\varepsilon_L$ being independent. We define $\eta_L = \sigma^2_{\varepsilon,L} / \sigma^2_L$.
The posterior distribution of the LFGP $Y_L$ given $ \textbf{Z}_L(X_{tr}^L)$ is the one of a GP 
$\widetilde{Y}_L$ with mean $m_{Y_L}(\textbf{x})$ and covariance $v_{Y_L}(\textbf{x},\textbf{x}')$ given by:
\[ 
\begin{cases}    
m_{Y_L}(\textbf{x}) = \textbf{f}_L(\textbf{x})^\top \boldsymbol\beta_L + \textbf{r}_L(\textbf{x};\boldsymbol\theta_L)^\top (R_L(\boldsymbol\theta_L) + \eta_L I_{N_L})^{-1}(\textbf{z}_{tr}^L - F_L\boldsymbol\beta_L) \\
v_{Y_L}(\textbf{x},\textbf{x}') = \sigma^2_L \left(r_L(\textbf{x},\textbf{x}';\boldsymbol\theta_L) - \textbf{r}_L(\textbf{x}; \boldsymbol\theta_L)^\top (R_L(\boldsymbol\theta_L) + \eta_L I_{N_L})^{-1}\textbf{r}_L(\textbf{x}';\boldsymbol\theta_L) \right)\!. 
\end{cases} 
\]
The LF parameters are estimated by maximizing the likelihood of the LF data:
\[ 
\begin{cases}
\hat{\boldsymbol\beta}_L = (F_L^{\top} (R_L(\boldsymbol\theta_L)+\eta_L I_{N_L})^{-1}F_L)^{-1} F_L^{\top}(R_L(\boldsymbol\theta_L)+\eta_L I_{N_L})^{-1}\textbf{z}_{tr}^L \\
\hat{\sigma}^2_L = (\textbf{z}_{tr}^L - F_L\hat{\boldsymbol\beta}_L)^{\top}(R_L(\boldsymbol\theta_L)+\eta_L I_{N_L})^{-1}(\textbf{z}_{tr}^L - F_L\hat{\boldsymbol\beta}_L) \, / \, N_L \\
(\hat{\boldsymbol\theta}_L, \hat{\eta}_L) =\argmin_{(\boldsymbol\theta_L, \, \eta_L)} \left[ N_L \log \!\left(\hat{\sigma}^{2}_L \right) + \log \det (R_L(\boldsymbol\theta_L)+\eta_L I_{N_L}) \right]\!.
\end{cases}
\]
The prior model for the HF function $y_H$ is a GP $Y_H$ defined as: 
\begin{equation}\label{eq:recursive_AR1}
\begin{cases}
Y_H(\textbf{x}) = \rho(\textbf{x})\widetilde{Y}_L(\textbf{x}) + \Delta_H(\textbf{x}) \\
\widetilde{Y}_L \perp \Delta_H, 
\end{cases}
\end{equation}
where $\widetilde{Y}_L \perp \Delta_H$ means that $\widetilde{Y}_L$ and $\Delta_H$ are independent. 
The GP $\Delta_H$ with mean $m_H(\textbf{x}) = \textbf{f}_H(\textbf{x})^\top \boldsymbol\beta_H$ and covariance $k_H(\textbf{x},\textbf{x}')=\sigma^2_H r_H(\textbf{x},\textbf{x}';\boldsymbol\theta_H)$ models the discrepancy.
Hence, $Y_H$ is a linear transformation of the posterior LF process $\widetilde{Y}_L$, whose parameters are already estimated, and not of the prior process $Y_L$ as in the non-recursive formulation (\cite{kennedy2000ohagan, Forrester2007function}).
We let $Z_H = Y_H + \varepsilon_H$, with $Y_H$ and $\varepsilon_H$ independent. We define $\eta_H = \sigma^2_{\varepsilon,H} / \sigma^2_H$ and $\mathrm{D}_\rho = \text{Diag}(\boldsymbol\rho(X^H_{tr}))$. \\

Our aim is to find the posterior distribution of the HF process $Y_H$ conditioned  by $\textbf{Z}_H(X_{tr}^H)=\textbf{z}_{tr}^H$ and to optimize the HF parameters. Both goals rely on the distribution of the Gaussian vector $\textbf{Z}_H(X_{tr}^H)$, which depends on the vector $\widetilde{\mathbf{Y}}\!_{L}(X_{tr}^H)$. If the experimental designs are nested, i.e. $X_{tr}^H \subset X_{tr}^L$, and if there is no LF noise so that the interpolation property is verified at the LF level, meaning that $\widetilde{\mathbf{Y}}\!_{L}(X_{tr}^L) = \textbf{z}_{tr}^L$ almost surely, then  $\widetilde{\mathbf{Y}}\!_{L}(X_{tr}^H) = \textbf{z}_L(X_{tr}^H)$ almost surely, where $\textbf{z}_L(X_{tr}^H)$ is the sub-vector of $\textbf{z}_{tr}^L$ corresponding to the inputs $X_{tr}^H$. In particular, $\widetilde{\mathbf{Y}}\!_{L}(X_{tr}^H)$ is of variance zero because it is almost surely equal to a deterministic quantity. This implies a simplification of the covariance matrix of the vector $\textbf{Z}_H(X_{tr}^H)$ given $\textbf{z}_{tr}^L$: 
\begin{equation}\label{eq:simple_HFcov}
   \textbf{Z}_H(X_{tr}^H) \sim \mathcal{N}\!\left( \mathrm{D}_\rho \textbf{z}_L(X^H_{tr}) + F_H \boldsymbol\beta_H, \sigma^2_H (R_H(\boldsymbol\theta_H) + \eta_H I_{N_H})\right)\!. 
\end{equation}
However, this simplification does not hold if LF noise is present or if the experimental designs are not nested. We define the auto-regressive mean and covariance functions, as well as their corresponding vectorized notations, when evaluated on the HF training data: 
\[  
\begin{cases}
m_{AR}(\textbf{x}) = \rho(\textbf{x})m_{Y_L}(\textbf{x}) +  \textbf{f}_H(\textbf{x})^\top \boldsymbol\beta_H \\
\textbf{m}_{AR} =  \mathrm{D}_\rho \textbf{m}_{Y_L}(X_{tr}^H) +  F_H\boldsymbol\beta_H \\
k_{AR}(\textbf{x},\textbf{x}') = \rho(\textbf{x})\rho(\textbf{x}') v_{Y_L}(\textbf{x},\textbf{x}') + \sigma^2_H r_H(\textbf{x},\textbf{x}';\boldsymbol\theta_H) \\
\textbf{k}_{AR}(\textbf{x}) = \rho(\textbf{x})\mathrm{D}_\rho \textbf{v}_{Y_L}(X_{tr}^H, \textbf{x}) + \sigma^2_H \textbf{r}_H(\textbf{x}; \boldsymbol\theta_H) \\
K_{AR} = \mathrm{D}_\rho V_{Y_L}(X^H_{tr},X^H_{tr}) \mathrm{D}_\rho + \sigma^2_H R_H(\boldsymbol\theta_H). 
\end{cases}
\]
The posterior distribution of the HFGP $Y_H$ given $\textbf{Z}_H(X_{tr}^H)=\textbf{z}_{tr}^H$ is the one of a GP $\widetilde{Y}_H$ with mean $m_{Y_H}$ and covariance $v_{Y_H}$ functions
obtained by applying the Gaussian conditioning theorem which yields the general co-kriging equations (\cite{legratiet_thesis}):
\[  
\begin{cases}    
m_{Y_H}(\textbf{x}) = m_{AR}(\textbf{x}) + \textbf{k}_{AR}(\textbf{x})^\top (K_{AR} + \sigma^2_{\varepsilon,H} I_{N_H})^{-1}(\textbf{z}_{tr}^H - \textbf{m}_{AR})  \\
v_{Y_H}(\textbf{x},\textbf{x}') = k_{AR}(\textbf{x},\textbf{x}') - \textbf{k}_{AR}(\textbf{x})^\top (K_{AR} + \sigma^2_{\varepsilon,H} I_{N_H})^{-1}\textbf{k}_{AR}(\textbf{x}').
\end{cases} 
\]
The previous equations are valid for any function $\rho$. We suppose from now on that the scaling term $\rho$ is a linear predictor function, i.e. that $\rho(\textbf{x}) = \textbf{g}_L(\textbf{x})^\top \boldsymbol\beta_\rho$, where $\boldsymbol\beta_\rho \in \mathbb{R}^{q \times 1}$ is a vector of parameters to estimate. We let as well the matrix $G_L = (g_{L,j}(\textbf{x}_i^H))_{(i,j) \in \{1,\dots, N_H\} \times\{1,\dots, q\}} \in \mathbb{R}^{N_H \times q}$, so that $\boldsymbol\rho(X_{tr}^H)=G_L\boldsymbol\beta_\rho$. We are now interested in maximizing the likelihood of $\textbf{Z}_H(X_{tr}^H)$ with respect to the HF parameters $\boldsymbol\xi_H =(\boldsymbol\beta_H, \boldsymbol\beta_\rho, \sigma^2_H, \boldsymbol\theta_H, \eta_H)$. The key point that prevents the previously mentioned simplification is that, when there is LF noise or when the input design sets are not nested, the vector $\widetilde{\mathbf{Y}}\!_{L}(X_{tr}^H)$ is not directly accessible and serves as a latent variable. Indeed, when $\widetilde{\mathbf{Y}}\!_{L}(X_{tr}^H)$ is observed, we have an expression similar to (\ref{eq:simple_HFcov}):
\[ 
\left( \textbf{Z}_H(X_{tr}^H)
\, | \, \widetilde{\mathbf{Y}}\!_{L}(X_{tr}^H) \right) \sim \mathcal{N}\!\left( \mathrm{D}_\rho \widetilde{\mathbf{Y}}\!_{L}(X_{tr}^H) + F_H \boldsymbol\beta_H, \sigma_H^2 (R_H(\boldsymbol \theta_H) + \eta_H I_{N_H})\right)\!.  
\]
The natural approach proposed by \citet[Section~2.4]{zertuche_thesis} uses the EM algorithm (\cite{Dempster1977EM}) in the deterministic, non-nested case. We show that this method can be used in the general case, when the output data is noisy for both fidelity levels. The version of the EM algorithm given in \cite{zertuche_thesis} is computationally heavy since all LF and HF parameters are optimized at once. By construction of the decoupled optimization, the LF parameters denoted by $\boldsymbol\xi_L =(\boldsymbol\beta_L, \sigma^2_L, \boldsymbol\theta_L, \eta_L)$ are considered fixed. We focus here on solely estimating the HF parameters. The goal of the procedure is to obtain estimates $(\bm{\xi}_H^{(t)})_{t \geq 1}$ such that the HF marginal log-likelihood $\ell_H(\bm{\xi}_H^{(t)}) = \log p (\textbf{z}_{tr}^H \, | \, \bm{\xi}_H^{(t)}, \bm{\xi}_L)$ is non-decreasing in $t$. The EM Q-function writes:
\begin{align*}
Q(\bm{\xi}_H \, | \, \bm{\xi}_H^{(t)}) &= \mathbb{E}_{( \widetilde{\mathbf{Y}}\!_{L}(X_{tr}^H)
 \, | \,\textbf{Z}_H(X_{tr}^H) =\textbf{z}_{tr}^H; \boldsymbol\xi_H^{(t)})}\left[\log p(\widetilde{\mathbf{Y}}\!_{L}(X_{tr}^H),\textbf{z}_{tr}^H \, | \,  \bm{\xi}_H, \bm{\xi}_L) \right] \\
 &= Q_{alt}(\bm{\xi}_H \, | \, \bm{\xi}_H^{(t)}) + \Omega(\bm{\xi}_H^{(t)}),
 \end{align*} where:
\[  
\begin{cases}
Q_{alt}(\bm{\xi}_H \, | \, \bm{\xi}_H^{(t)}) = \mathbb{E}_{( \widetilde{\mathbf{Y}}\!_{L}(X_{tr}^H)
\, | \,\textbf{Z}_H(X_{tr}^H) = \textbf{z}_{tr}^H; \boldsymbol\xi_H^{(t)})}\left[\log p(   \textbf{z}_{tr}^H \, | \, \widetilde{\mathbf{Y}}\!_{L}(X_{tr}^H), \bm{\xi}_H, \bm{\xi}_L) \right]  \\[3mm]
\Omega(\bm{\xi}_H^{(t)}) = \mathbb{E}_{( \widetilde{\mathbf{Y}}\!_{L}(X_{tr}^H)
\, | \,\textbf{Z}_H(X_{tr}^H) = \textbf{z}_{tr}^H; \boldsymbol\xi_H^{(t)})}\left[\log p( \widetilde{\mathbf{Y}}\!_{L}(X_{tr}^H) \, | \, \bm{\xi}_L) \right]\!.
\end{cases}
\]
Since $\Omega$ does not depend on $\bm{\xi}_H$, we describe the EM algorithm using the alternative Q-function $Q_{alt}$, as it is a valid substitute for $Q$ in the sense that increasing $Q_{alt}$ during the EM iterations results in increasing $\ell_H$. Guarantees regarding the monotonicity of $\ell_H$ will be stated clearly afterwards. In the following, we denote $\textbf{1}_N$ the column vector of size $N$ composed of only ones, $\mathbf{0}_{M \times N}$ is the null matrix of size $M \times N$, $\boldsymbol\beta_{\rho,H} = (\boldsymbol\beta_\rho^\top, \boldsymbol\beta_H^\top)^\top$ is the column vector that concatenates $\boldsymbol\beta_\rho$ and $\boldsymbol\beta_H$ and $\mathcal{S}$ is the quadratic form in $\textbf{x}$: $\mathcal{S}(\textbf{x}; \Sigma)= \textbf{x}^\top \Sigma^{-1}\textbf{x}$, with $\textbf{x}$ a column vector and $\Sigma$ a SPD matrix. In our case, the steps can be written as: 
\\

\textbf{E-step:} Using the values $\bm{\xi}_H^{(t)}$ of the HF parameters at iteration $t$, compute the following vectors and matrices:
\begin{equation}\label{eq:Many_Matrices}
\begin{cases}
\Sigma_{\textbf{YZ}}^{(t)} =  V_{Y_L}(X_{tr}^H,X_{tr}^H) \mathrm{D}_\rho^{(t)} \\
\Sigma_{\textbf{ZZ}}^{(t)} = \mathrm{D}_\rho^{(t)} V_{Y_L}(X^H_{tr},X^H_{tr} ) \mathrm{D}_\rho^{(t)}  + \sigma^{2 (t)}_H \!\left(R_H(\boldsymbol\theta_H^{(t)}) + \eta_H^{(t)}I_{N_H} \right) \\
\boldsymbol\mu_{\textbf{Y}|\textbf{Z}}^{(t)} = \textbf{m}_{Y_L}(X_{tr}^H) + \Sigma_{\textbf{YZ}}^{(t)}(\Sigma_{\textbf{ZZ}}^{(t)})^{-1} \!\left(\textbf{z}_{tr}^H - \mathrm{D}_\rho^{(t)} \textbf{m}_{Y_L}(X_{tr}^H) -  F_H\boldsymbol\beta_H^{(t)} \right)  \\
\Sigma_{\textbf{Y}|\textbf{Z}}^{(t)} = V_{Y_L}(X^H_{tr},X^H_{tr}) - \Sigma_{\textbf{YZ}}^{(t)}(\Sigma_{\textbf{ZZ}}^{(t)})^{-1}(\Sigma_{\textbf{YZ}}^{(t)})^\top \\
\mathcal{T}^{(t)}(\boldsymbol\theta_H,\eta_H) = G_L^\top \!\left((R_H(\boldsymbol\theta_H) + \eta_H I_{N_H})^{-1} \odot \Sigma_{\textbf{Y}|\textbf{Z}}^{(t)} \right) \!G_L,
\end{cases}
\end{equation} 
where the symbol $\odot$ corresponds to the element-wise product, also called the Hadamard product. We also introduce the block matrices:
\begin{equation}\label{eq:EM_blockmatrices}
H_H^{(t)} = \left( G_L \odot (\boldsymbol\mu_{\textbf{Y}|\textbf{Z}}^{(t)}\textbf{1}_q^\top), \,\, F_H \right) \, \, \,  \text{and} \, \, \,  T^{(t)}(\boldsymbol\theta_H,\eta_H) =  \begin{pmatrix}
\mathcal{T}^{(t)}(\boldsymbol\theta_H,\eta_H)  & \mathbf{0}_{q \times p_H}  \\
\mathbf{0}_{p_H \times q} &  \mathbf{0}_{p_H \times p_H}
\end{pmatrix}.  \end{equation}
Finally, we compute the alternative $Q$-function:
\begin{align*}
Q_{alt}(\bm{\xi}_H;\bm{\xi}_H^{(t)}) &=  -\frac{N_H}{2}\log(\sigma_H^2) - \frac{1}{2}\log \det (R_H(\boldsymbol\theta_H) + \eta_H I_{N_H}) - \frac{1}{2\sigma_H^2} \boldsymbol\beta_{\rho,H}^\top {T}^{(t)}(\boldsymbol\theta_H,\eta_H) \boldsymbol\beta_{\rho,H}\\
&- \frac{1}{2\sigma_H^2} \mathcal{S}\!\left(\textbf{z}_{tr}^H -H_H^{(t)}\boldsymbol\beta_{\rho,H}; R_H(\boldsymbol\theta_H) + \eta_H I_{N_H}\right)-\frac{N_H}{2}\log(2\pi). 
\end{align*}
The details of the algebraic derivations are available in Appendix \ref{app:EM_details}. \\

\textbf{M-step:} We are now interested in the maximization of $Q_{alt}$ with respect to the HF parameters $\bm{\xi}_H$. Proceeding in a similar manner as for the LF parameter estimation, we first derive a formula for the updated value of $\boldsymbol\beta_{\rho, H}$, which is possible since we managed to express $Q_{alt}$ as a quadratic form in $\boldsymbol\beta_{\rho, H}$:
\[ 
\hat{\boldsymbol\beta}_{\rho,H}^{(t+1)} = \left( H_H^{(t)\top}(R_H(\boldsymbol \theta_H) + \eta_H I_{N_H})^{-1} H_H^{(t)}  + T^{(t)}(\boldsymbol\theta_H,\eta_H)\right)^{-1} \!H_H^{(t)\top}(R_H(\boldsymbol \theta_H) + \eta_H I_{N_H})^{-1}\textbf{z}_{tr}^H, 
\]
replacing $\boldsymbol\beta_{\rho,H}$ by the previous formula and maximizing $Q_{alt}$ with respect to $\sigma_H^2$ gives: 
\[
\hat{\sigma}^{2 (t+1)}_H = \frac{1}{N_H}\left[ \mathcal{S}\! \left(\textbf{z}_{tr}^H -H_H^{(t)}\hat{\boldsymbol\beta}_{\rho,H}^{(t+1)}; R_H(\boldsymbol \theta_H) + \eta_H I_{N_H} \right) + \hat{\boldsymbol\beta}_{\rho,H}^{(t+1)\top}T^{(t)}(\boldsymbol\theta_H,\eta_H)\hat{\boldsymbol\beta}_{\rho,H}^{(t+1)}  \right]\!.
\]
We introduce the profiled Q-function:
\begin{align*}
\widetilde{Q}_{alt}\!\left((\boldsymbol\theta_H,\eta_H); \bm{\xi}_H^{(t)} \right) &= Q_{alt}  \!\left((\hat{\boldsymbol\beta}_{\rho,H}^{(t+1)}, \hat{\sigma}^{2 (t+1)}_H, \boldsymbol\theta_H,\eta_H);\bm{\xi}_H^{(t)} \right) \\
&= -\frac{N_H}{2} \log \!\left(\hat{\sigma}^{2 (t+1)}_H \right) -\frac{1}{2} \log \det  (R_H(\boldsymbol \theta_H) + \eta_H I_{N_H}) -\frac{N_H}{2} (1+\log(2\pi)), 
\end{align*}
and the remaining parameters must be optimized numerically: 
\[ 
\left(\boldsymbol\theta_H^{(t+1)}, \eta_H^{(t+1)} \right)  = \argmax_{\boldsymbol\theta_H, \, \eta_H} \,\,\widetilde{Q}_{alt}\!\left((\boldsymbol\theta_H,\eta_H); \bm{\xi}_H^{(t)}\right)\!.
\]
These results are analogous to those of \cite{zertuche_thesis}. One important difference is that we do not suppose that $\rho$ is a known function. We managed to generalize the result of \cite{zertuche_thesis} in the case where the observations are noisy and by recovering an update formula for $\boldsymbol\beta_\rho$ at the M-step when $\rho$ writes $\rho(\textbf{x}) = \textbf{g}_L(\textbf{x})^\top\boldsymbol\beta_\rho$. When $\Sigma_{\textbf{Y}|\textbf{Z}}^{(t)} = 0$, i.e. when $\widetilde{\mathbf{Y}}\!_{L}(X_{tr}^H)$ is observed, $T^{(t)}$ vanishes and the M-step reduces to the estimator of \cite{legratiet2014rec_cokriging} in the nested case without LF noise. We now address the monotonicity of $\ell_H$ with the following proposition:
\begin{prop}\label{prop:EM_mono}
Let $\boldsymbol\xi_H^{(t)} =(\boldsymbol\beta_{\rho,H}^{(t)}, \sigma^{2 (t)}_H, \boldsymbol\theta_H^{(t)}, \eta_H^{(t)})$ and let $(\boldsymbol\theta_H^{(t+1)}, \eta_H^{(t+1)})$ such that:
\begin{equation}\label{eq:GEM}
\widetilde{Q}_{alt}\!\left((\boldsymbol\theta_H^{(t+1)},\eta_H^{(t+1)}); \bm{\xi}_H^{(t)}\right) \geq \widetilde{Q}_{alt}\!\left((\boldsymbol\theta_H^{(t)},\eta_H^{(t)}); \bm{\xi}_H^{(t)} \right)\!.
\end{equation}
Set  $\boldsymbol\xi_H^{(t+1)} =(\hat{\boldsymbol\beta}_{\rho,H}^{(t+1)}, \hat{\sigma}^{2 (t+1)}_H, \boldsymbol\theta_H^{(t+1)}, \eta_H^{(t+1)})$, then $\ell_H(\bm{\xi}_H^{(t+1)}) \geq  \ell_H(\bm{\xi}_H^{(t)})$.
\end{prop}
\paragraph{Proof:}
By definition of $\widetilde{Q}_{alt}$ and $\boldsymbol\xi_H^{(t+1)}$, and by (\ref{eq:GEM}):
\[   
Q_{alt}(\bm{\xi}_H^{(t+1)};\bm{\xi}_H^{(t)}) = \widetilde{Q}_{alt}\!\left((\boldsymbol\theta_H^{(t+1)},\eta_H^{(t+1)}); \bm{\xi}_H^{(t)}\right)
\geq
\widetilde{Q}_{alt}\!\left((\boldsymbol\theta_H^{(t)},\eta_H^{(t)}); \bm{\xi}_H^{(t)}\right)
\geq 
Q_{alt}(\bm{\xi}_H^{(t)};\bm{\xi}_H^{(t)}). 
\]
We recall that $Q(\bm{\xi}_H \, | \, \bm{\xi}_H^{(t)}) = Q_{alt}(\bm{\xi}_H \, | \, \bm{\xi}_H^{(t)}) + \Omega(\bm{\xi}_H^{(t)})$, and one can write the log-likelihood as: $   \ell_H(\bm{\xi}_H) = Q(\bm{\xi}_H \, | \, \bm{\xi}_H^{(t)}) - \mathbb{H}(\bm{\xi}_H \, | \, \bm{\xi}_H^{(t)})$ with:
\[
\mathbb{H}(\bm{\xi}_H \, | \, \bm{\xi}_H^{(t)}) = \mathbb{E}_{( \widetilde{\mathbf{Y}}\!_{L}(X_{tr}^H)
\, | \,\textbf{Z}_H(X_{tr}^H) = \textbf{z}_{tr}^H; \boldsymbol\xi_H^{(t)})}\left[\log p(    \widetilde{\mathbf{Y}}\!_{L}(X_{tr}^H) \, | \,\textbf{z}_{tr}^H , \bm{\xi}_H, \bm{\xi}_L) \right]. 
\]
Gibbs' inequality yields $\mathbb{H}(\bm{\xi}_H^{(t+1)} \, | \, \bm{\xi}_H^{(t)}) \leq \mathbb{H}(\bm{\xi}_H^{(t)} \, | \, \bm{\xi}_H^{(t)})$ (\cite{Dempster1977EM}), therefore: 
\begin{align*}
\ell_H(\bm{\xi}_H^{(t+1)}) - \ell_H(\bm{\xi}_H^{(t)})
&\geq Q(\bm{\xi}_H^{(t+1)} \mid \bm{\xi}_H^{(t)}) - Q(\bm{\xi}_H^{(t)} \mid \bm{\xi}_H^{(t)}) \\
&= Q_{alt}(\bm{\xi}_H^{(t+1)} \mid \bm{\xi}_H^{(t)}) - Q_{alt}(\bm{\xi}_H^{(t)} \mid \bm{\xi}_H^{(t)}) \geq 0. \tag*{$\square$}
\end{align*} 

We use a multi-start strategy for each instance of numerical optimization in the algorithm, in order to mitigate the influence of the initialization. Condition (\ref{eq:GEM}) is enforced in each iteration of the EM, i.e. if none of the multi-start solutions increases $\widetilde{Q}_{alt}$, the current point is kept: $(\boldsymbol\theta_H^{(t+1)},\eta_H^{(t+1)}) = (\boldsymbol\theta_H^{(t)},\eta_H^{(t)})$. As for the LF level optimization, we use the L-BFGS-B algorithm (\cite{zhu1997L_BFGS_B}) which exploits the gradient of $\widetilde{Q}_{alt}$, its expression is given in Appendix \ref{app:EM_details}. Note that $\ell_H$ is available in closed form, since $\textbf{Z}_H(X_{tr}^H) \sim \mathcal{N}(\mathbf{m}_{AR}, K_{AR} + \sigma^2_{\varepsilon,H} I_{N_H})$. It could therefore be maximized directly with respect to $(\boldsymbol\beta_\rho, \sigma^2_H, \boldsymbol\theta_H, \eta_H)$, $\boldsymbol\beta_H$ being profiled out. We rather use the EM algorithm because each M-step has the same structure as single-fidelity MLE with several closed-form update formulas. The EM algorithm is not guaranteed to be faster than a direct maximization of $\ell_H$, which is in dimension $D+2+q$, where $D$ is the input dimension and $q$ is the size of $\beta_\rho$. However, in the regime $N_L \gg N_H$, which is the one of practical interest because LF data is cheap, the training time is dominated by the fit of the LFGP (Tables \ref{tab:sine_EM_time} and \ref{tab:park_EM_time} in Section \ref{sec:Num_applications}). The choice of the HF optimizer has a limited impact on the total cost in this case.

\subsection{Non-recursive formulation and fully coupled optimization}

We recall in this subsection the non-recursive formulation proposed by \cite{kennedy2000ohagan}. In this formulation, the prior HFGP is defined using the prior LFGP $Y_L$ and not the posterior $\widetilde{Y}_L$: $Y_H^{KO}(\textbf{x}) = \rho(\textbf{x})Y_L(\textbf{x}) + \Delta_H(\textbf{x})$. The GPs $Y_L$ and $\Delta_H$ have the same distributions as in the previous subsection, $(Y_L,\Delta_H, \varepsilon_L, \varepsilon_H)$ are also mutually independent. We define $\textbf{z}_{LH}^\top = ((\textbf{z}_{tr}^L)^\top,(\textbf{z}_{tr}^H)^\top)$, $\boldsymbol\beta_{LH}^\top=(\boldsymbol\beta_L^\top, \boldsymbol\beta_H^\top)$, $\textbf{f}_{LH}(\textbf{x})^\top=(\rho(\textbf{x}) \textbf{f}_L(\textbf{x})^\top, \textbf{f}_H(\textbf{x})^\top)$ and: 
\[   
F_{LH} = \begin{pmatrix}
F_L & \mathbf{0}_{N_L \times p_H} \\
\mathrm{D}_\rho f_L(X_{tr}^H) & F_H
\end{pmatrix}\! \quad \text{with} \quad f_L(X_{tr}^H) =  (f_{L,j}(\textbf{x}_i^H))_{(i,j) \in \{1,\dots, N_H\} \times\{1,\dots, p_L\}}, 
\]
\[ 
\mathbf{v}_{LH}(\textbf{x})^\top= \left(\rho(\textbf{x}) \sigma^2_L \textbf{r}_L(\textbf{x}; \boldsymbol\theta_L)^\top, \rho(\textbf{x})  \sigma_L^2 \textbf{r}_L(\textbf{x},X_{tr}^H; \boldsymbol\theta_L) \mathrm{D}_\rho + \sigma^2_H \textbf{r}_H(\textbf{x}; \boldsymbol\theta_H)^\top \right)\!.  
\]
The coupled LF-HF covariance matrix takes the form:
\[  
\widetilde{V}_{LH} = \begin{pmatrix}
\sigma_L^2 R_L(\boldsymbol\theta_L) + \sigma^2_{\varepsilon,L} I_{N_L} &  \sigma_L^2 r_L(X_{tr}^L, X_{tr}^H; \boldsymbol\theta_L)\mathrm{D}_\rho \\
\mathrm{D}_\rho\sigma_L^2 r_L(X_{tr}^H, X_{tr}^L; \boldsymbol\theta_L) &  \mathrm{D}_\rho \sigma^2_L r_L(X_{tr}^H, X_{tr}^H; \boldsymbol\theta_L) \mathrm{D}_\rho+ \sigma^2_H R_H(\boldsymbol\theta_H) +\sigma^2_{\varepsilon,H} I_{N_H}
\end{pmatrix}\!.
\]
The process $Y_H^{KO}$ conditioned by all data samples $(\textbf{Z}_L(X_{tr}^L)=\textbf{z}_{tr}^L,\textbf{Z}_H(X_{tr}^H)=\textbf{z}_{tr}^H)$ is a GP with the following mean and covariance functions:
\[ 
\begin{cases}
m_{Y_H^{KO}}(\textbf{x}) = \textbf{f}_{LH}(\textbf{x})^\top \boldsymbol\beta_{LH} + \mathbf{v}_{LH}(\textbf{x})^\top \widetilde{V}_{LH}^{-1}(\textbf{z}_{LH} - F_{LH}\boldsymbol\beta_{LH}) \\
v_{Y_H^{KO}}(\textbf{x}, \textbf{x}') = \rho(\textbf{x}) \rho(\textbf{x}')\sigma_L^2 r_L(\textbf{x},\textbf{x}';\boldsymbol\theta_L) + \sigma_H^2 r_H(\textbf{x},\textbf{x}';\boldsymbol\theta_H) - \mathbf{v}_{LH}(\textbf{x})^\top \widetilde{V}_{LH}^{-1}\mathbf{v}_{LH}(\textbf{x}').
\end{cases}
\]
\cite{legratiet2014rec_cokriging} have shown that the previous functions are equal respectively to the recursive mean $m_{Y_H}$ and covariance function $v_{Y_H}$ in the noise-free and nested case. This is also true in the general case, as stated by the following theorem:
\begin{thm}\label{thm:equiv_rec_nrec}
With all parameters fixed, we have for all $\mathbf{x},\mathbf{x}' \in \mathbb{R}^D$:
\[   
\begin{cases}
m_{Y_H}(\mathbf{x})=m_{Y_H^{KO}}(\mathbf{x}) \\
v_{Y_H}(\mathbf{x},\mathbf{x}')=v_{Y_H^{KO}}(\mathbf{x},\mathbf{x}').
\end{cases}
\]
\end{thm} 
The proof is given in Appendix \ref{app:proof_thm}. As in simple-fidelity GP regression, one can derive a MLE formula for the mean parameters:
\[ 
\hat{\boldsymbol\beta}_{LH} = \left( F_{LH}^{\top} \widetilde{V}_{LH}^{-1}F_{LH} \right)^{-1} F_{LH}^{\top}\widetilde{V}_{LH}^{-1}\textbf{z}_{LH}, 
\]
the fully coupled log-likelihood takes the form:
\begin{align*}
\log \mathcal{L}_{LH}(\textbf{z}_{LH} \, | \,  \beta_{\rho}, \sigma^2_L, \sigma^2_H, \boldsymbol\theta_L, \boldsymbol\theta_H, \sigma^2_{\varepsilon,L}, \sigma^2_{\varepsilon,H}) &= -\frac{N_L+N_H}{2}\log(2\pi) - \frac{1}{2}\log \det \widetilde{V}_{LH} \\
& -\frac{1}{2} \!\left(\textbf{z}_{LH}-F_{LH}\hat{\boldsymbol\beta}_{LH} \right)^\top\! \widetilde{V}_{LH}^{-1} \!\left(\textbf{z}_{LH}-F_{LH}\hat{\boldsymbol\beta}_{LH} \right)
\end{align*}
and is jointly maximized with respect to all parameters using the L-BFGS-B algorithm with analytical gradients and multi-starts. A notable difference between the two approaches lies in the dimension of the numerical optimizations. We recall that $D$ is the dimension of the input space and the size of the correlation length vectors, $q$ is the size of $\beta_\rho$. The proposed approach addresses the bottleneck mentioned in \cite{Babaee2020CapeCod}: the fully coupled likelihood is optimized over a space of dimension $2(D+2)+q$, whereas in the decoupled EM-based approach, a sequence of optimization problems in dimension $D+1$ is solved. Theorem \ref{thm:equiv_rec_nrec} shows that the recursive and non-recursive formulations yield in fact the same predictive distribution for fixed parameters. Therefore, practical differences between the two arise solely from their specific parameter estimation strategy. The two formulations are compared in terms of wall-clock training time but also in terms of predictive performance of the resulting MFGP. The prediction cost after training is similar for both formulations. The following section introduces the relevant metrics.

\section{Performance metrics}\label{sec:UQ_metrics}

We present in this section several metrics in order to evaluate the performance of a surrogate model when the target function $y$ is known. We denote $X_{test} = (\textbf{x}_1,\dots,\textbf{x}_{N_{test}})$ the test input set, and $m$ and $s^2$ the posterior mean and variance functions of the latent function, given the observations and the plug-in estimation of the parameters. The predictivity coefficient denoted $Q^2$ is used to assess the performance of the mean prediction:
\[ 
Q^2 = 1-\frac{\sum_{t=1}^{N_{test}}  (y(\textbf{x}_t)-m(\textbf{x}_t))^2}{\sum_{t=1}^{N_{test}} (y(\textbf{x}_t) - \bar{y})^2} \quad \text{with} \quad \bar{y} = \frac{1}{{N_{test}}}\sum_{t=1}^{N_{test}} y(\textbf{x}_t). 
\]
The closer $Q^2$ is to $1$, the better the prediction, whereas $Q^2=0$ indicates that the surrogate model has the same prediction performance as the constant model that always outputs the empirical mean of the true values. Since $Q^2 \leq 1$, we often consider the quantity $1-Q^2 \geq 0$ for plots in logarithmic scale, it can be seen as a normalized mean squared error metric. The $Q^2$ metric does not give any information regarding the uncertainty on the predictions. Since the posterior distribution of the latent function at $\mathbf{x}$ is
$\mathcal{N}\big(m(\mathbf{x}), s(\mathbf{x})^2\big)$, the credible interval (CI) of level $\alpha \in (0,1)$ is:
\[ 
\mathcal{I}_\alpha(\textbf{x}) = \left[m(\textbf{x}) - \varphi_\alpha s(\textbf{x}),m(\textbf{x}) + \varphi_\alpha s(\textbf{x}) \right]\!, 
\]
where $\varphi_\alpha = \Phi^{-1}((1+\alpha)/2)$, $\Phi$ being the cumulative distribution function (CDF) of the $\mathcal{N}(0,1)$ distribution. This is an empirical Bayes CI due to the plug-in estimation of the parameters. One can also use the prediction interval (PI) on noisy outputs which is obtained by replacing $s(\textbf{x})$ by $\sqrt{s(\textbf{x})^2+ \hat{\sigma}^2_\varepsilon}$ in $\mathcal{I}_\alpha(\textbf{x})$:
\[ 
\mathcal{I}_{\alpha, \hat{\sigma}^2_\varepsilon}(\textbf{x}) = \left[m(\textbf{x}) - \varphi_\alpha \sqrt{s(\textbf{x})^2+ \hat{\sigma}^2_\varepsilon}, m(\textbf{x}) + \varphi_\alpha\sqrt{s(\textbf{x})^2+ \hat{\sigma}^2_\varepsilon} \right]\!, 
\]
where $\hat{\sigma}^2_\varepsilon$ is the estimated value for the noise variance. We define several metrics from these intervals. The mean empirical coverage probability (CP) is defined for CIs and PIs (\cite{ACHARKI2023}) as:
\[ 
\text{CICP}_{\alpha} = \displaystyle \frac{1}{N_{test}} \sum_{t=1}^{N_{test}}\! \mathbbm{1}\!\left(y(\textbf{x}_t) \in \mathcal{I}_\alpha(\textbf{x}_t) \right)
\quad \text{and} \quad
\text{PICP}_{\alpha} = \displaystyle  \frac{1}{N_{test}} \sum_{t=1}^{N_{test}}\! \mathbbm{1}\!\left(z_t \in \mathcal{I}_{\alpha, \hat{\sigma}^2_\varepsilon}(\textbf{x}_t)\right)
\]
where $z_t$ is a noisy observation at the test input $\textbf{x}_t$, whose noise is independent
of the training data. For the analytical test cases, $z_t = y(\textbf{x}_t) + \varepsilon_t$ is
simulated with $\varepsilon_t \sim \mathcal{N}(0, \sigma^2_\varepsilon)$, $\sigma^2_\varepsilon$
being the true noise variance. For real data, $z_t$ is the measurement at $\textbf{x}_t$ since $y$ is not available, and $Q^2$ is evaluated with the measurements as target values. We consider also the mean widths of CIs and PIs:
\[
\text{MCIW}_{\alpha} = \displaystyle \frac{1}{N_{test}} \sum_{t=1}^{N_{test}} 2 \varphi_\alpha s(\textbf{x}_t) \quad \text{and} \quad
\text{MPIW}_{\alpha} = \displaystyle  \frac{1}{N_{test}} \sum_{t=1}^{N_{test}} 2 \varphi_\alpha \sqrt{s(\textbf{x}_t)^2+ \hat{\sigma}^2_\varepsilon}. 
\]
The target value for both $\text{CICP}_{\alpha}$ and $\text{PICP}_{\alpha}$ is $\alpha$. In practice, instead of fixing a particular value for $\alpha$, it is advised to compare the values of $\text{CICP}_{\alpha}$ (resp. $\text{PICP}_{\alpha}$) against $\alpha$ for many values for $\alpha$ on $(0,1)$ in order to get the $\alpha$-CI plot (resp. $\alpha$-PI plot), as suggested by \cite{demay2022alphaCI}. For instance, if the values of $\text{CICP}_{\alpha}$ are above $\alpha$, it means that the GP variance is over-estimated. In other words, the surrogate model is under-confident. We say that the variance of the model is well calibrated, or that the intervals are reliable when the $\alpha$-CI and $\alpha$-PI plots are close to the diagonal $y=x$ line.
For repeated numerical experiments, we prefer a scalar metric rather than an entire plot for each computation. This is why we use the integral absolute error (IAE, \cite{marrel2024gps_insights}) defined as:
\[ 
\text{IAE}_{\text{CI}} = \int_0^1 |\text{CICP}_{\alpha} - \alpha  |\mathrm{d}\alpha \quad \text{and} \quad
\text{IAE}_{\text{PI}} = \int_0^1 |\text{PICP}_{\alpha} - \alpha |\mathrm{d}\alpha.
\]
The closer the IAE is to $0$, the more reliable the credible or prediction intervals are. The IAE metric summarizes the information contained in the $\alpha$-CI or $\alpha$-PI plot, but it does not, however, indicate if the model tends to overestimate or underestimate the predictive uncertainty. The previous integrals are computed using the trapezoidal rule on a uniform grid $\alpha \in \{0.001, 0.002, \dots, 0.999\}$.

\section{Numerical applications}\label{sec:Num_applications}

In the following, three application cases of increasing complexity are presented: an analytical 1D test function taken from \cite{giannoukou2025MFPCE}, the 4D Park function (\cite{Cox2001Parkfunction, Xiong2013test_fcts}) and a real-world case studied in \cite{Babaee2020CapeCod}. We use the same GP model each time, i.e. the prior parametric mean functions $m_L$ and $m_H$ and the scaling function $\rho$ are constants, and we take the Gaussian correlation function for the GP covariances: 
\[
k(\textbf{x},\textbf{x}';\boldsymbol\theta) = \sigma^2 \exp \!\left( - \frac{1}{2} \sum_{d=1}^D \left(\frac{x^{(d)}-x'^{(d)}}{\theta_d}\right)^2 \right)\!. 
\]
Hence, in every application, the following parameters are always scalars: $\beta_L$, $\beta_H$, $\beta_\rho$, $\sigma^2_L$, $\sigma^2_H$, as well as the noise parameters: $\eta_L$ and $\eta_H$ for the recursive approach and $\sigma^2_{\varepsilon,L}, \sigma^2_{\varepsilon,H}$ for the non-recursive approach. The size of the length scale vectors $\boldsymbol\theta_L$ and $\boldsymbol\theta_H$ corresponds to the dimension $D$ of the input $\textbf{x}$. The Gaussian kernel is justified here since the two analytical test functions are smooth and it is also used in \cite{Babaee2020CapeCod}. Further remarks about the implementation and the optimization are given in Appendix \ref{app:implementation}.

\subsection{Analytical 1D case}

We consider in this section a simple analytical one-dimensional function to which additive Gaussian noise is added (\cite{giannoukou2025MFPCE}). The LF and HF test functions are defined for $x \in [0,2]$ as: 
\[
\begin{cases}
y_L(x) = \sin(2\pi x) \\
y_H(x) = \left(\frac{x}{4} - \sqrt 2\right)  \sin(2 \pi x + \pi).
\end{cases}
\]
We consider several configurations of experiment designs. The number of training data points varies: $N_L \in \{ 100, 500, 1000 \}$ and $N_H \in \{5, 10, 20 \}$. The input data points are obtained through standard Latin hypercube sampling (LHS, \cite{McKay1979LHS}). The LF and HF input designs are sampled independently, thus they are almost surely not nested. The LF and HF output points are both tainted by artificial Gaussian noise with the respective standard deviations: $\sigma_{\varepsilon,L} = 0.3$ and $\sigma_{\varepsilon,H} = 0.1$. We replicate $N_{rep} = 100$ times each design in order to assess the variability of the metrics because of the randomness in the point selection and the noise realization. The test set is composed of $N_{test}=10^4$ points uniformly distributed in $[0,2]$. We also compare the performance of the MFGP AR(1) models presented earlier to a single-fidelity GP model trained only on the HF data, as a baseline. The single-fidelity GP model is referred to as the HF-only-GP and the recursive and non-recursive MFGPs are respectively called R-MFGP and NR-MFGP.  \\

\begin{figure}[H]
\centering
\includegraphics[height=6.5cm]{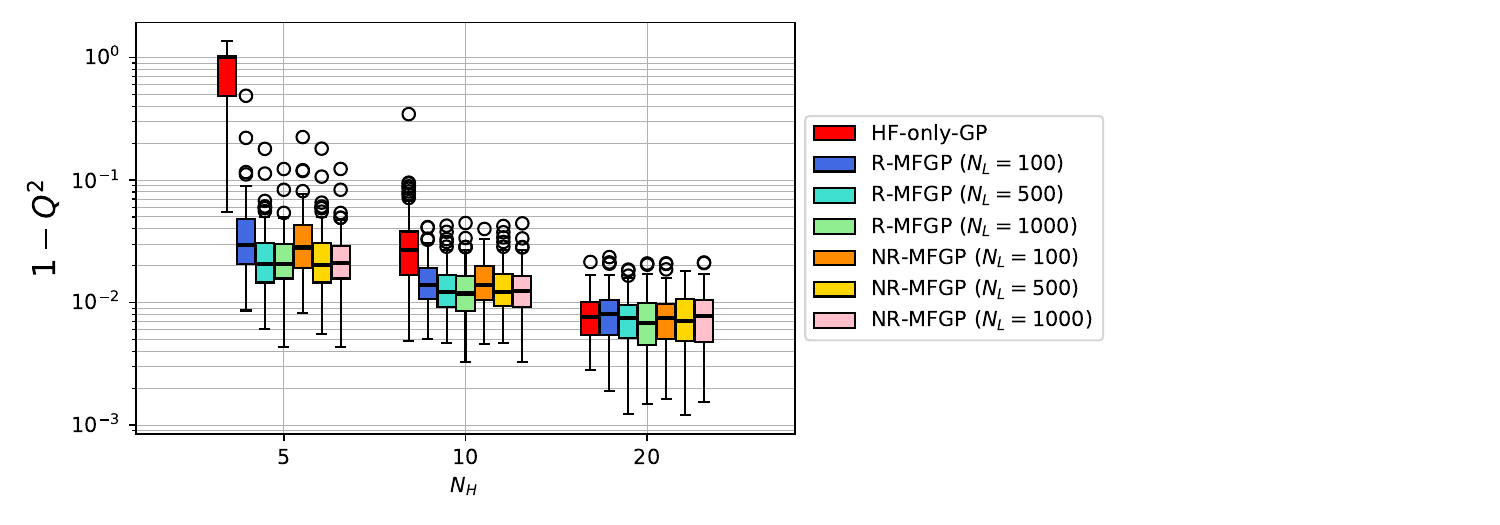}
\caption[]{Prediction error $1 - Q^2$ (log scale) of the HF-only-GP and of the recursive (R-MFGP) and non-recursive (NR-MFGP) formulations. Box-plots are computed over $100$ random designs.}
\label{fig:Errors_sine}
\end{figure}

\begin{figure}[H]
\centering
\includegraphics[height=6.5cm]{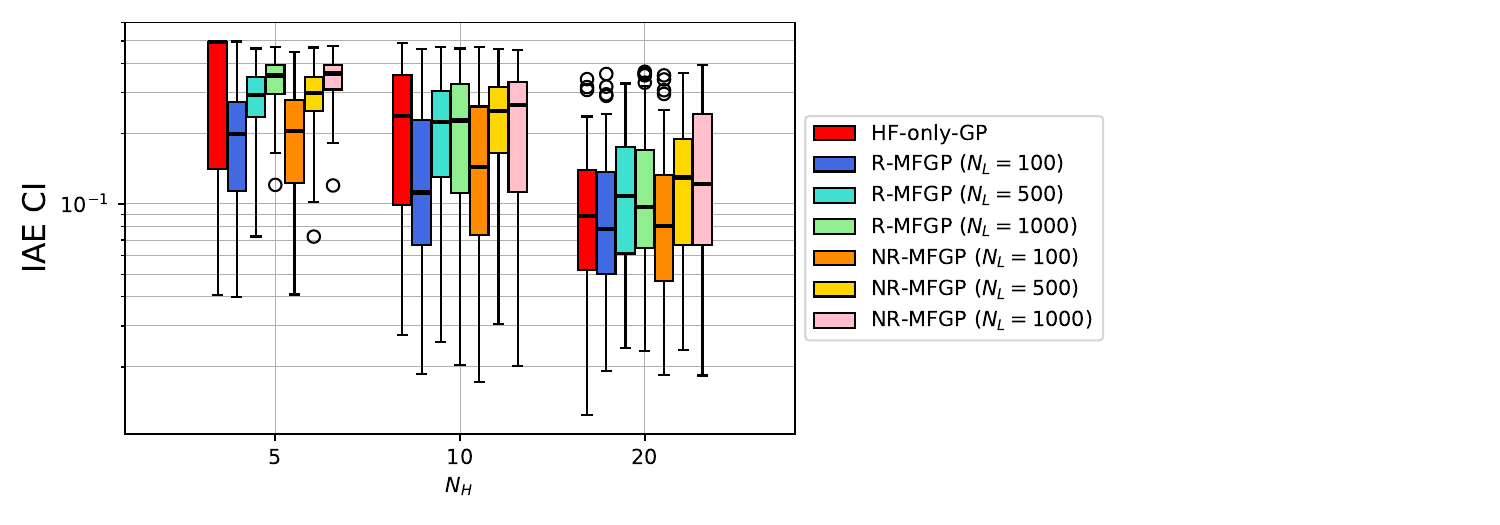}
\caption[]{$\text{IAE}_{\text{CI}}$ (log scale) of the HF-only-GP and of the recursive (R-MFGP) and non-recursive (NR-MFGP) formulations. Box-plots are computed over $100$ random designs.}
\label{fig:IAE_CI_sine}
\end{figure}

\begin{figure}[H]
\centering
\includegraphics[height=6.5cm]{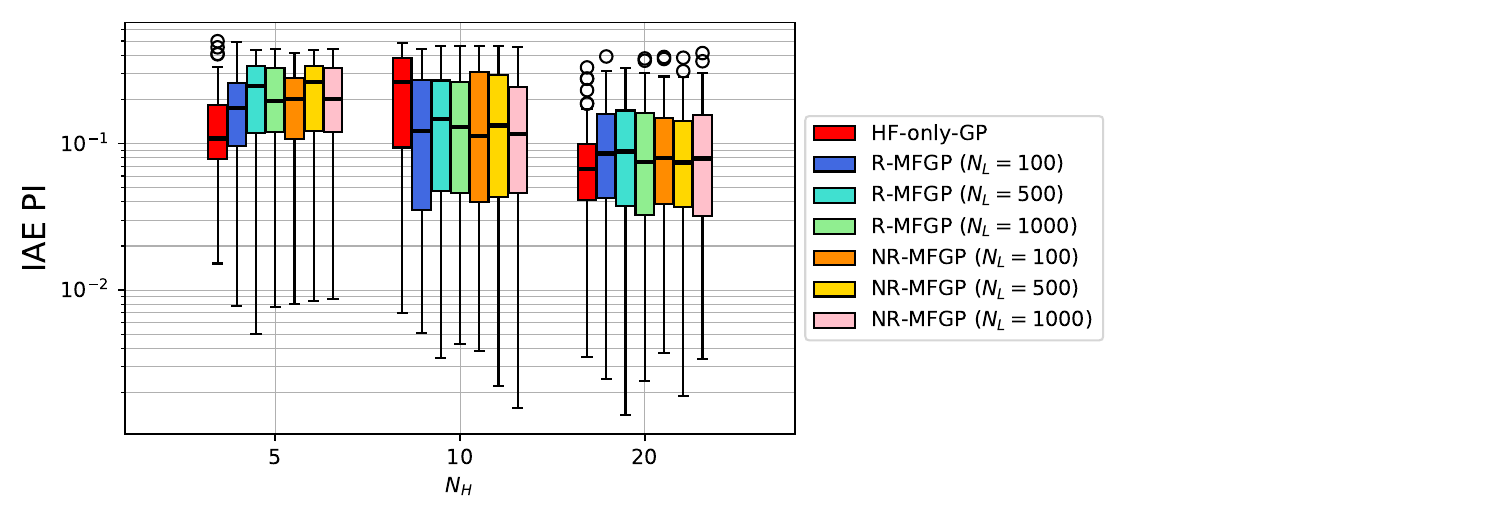}
\caption[]{$\text{IAE}_{\text{PI}}$ (log scale) of the HF-only-GP and of the recursive (R-MFGP) and non-recursive (NR-MFGP) formulations. Box-plots are computed over $100$ random designs.}
\label{fig:IAE_PI_sine}
\end{figure}

\begin{figure}[H]
\centering
\includegraphics[height=6.5cm]{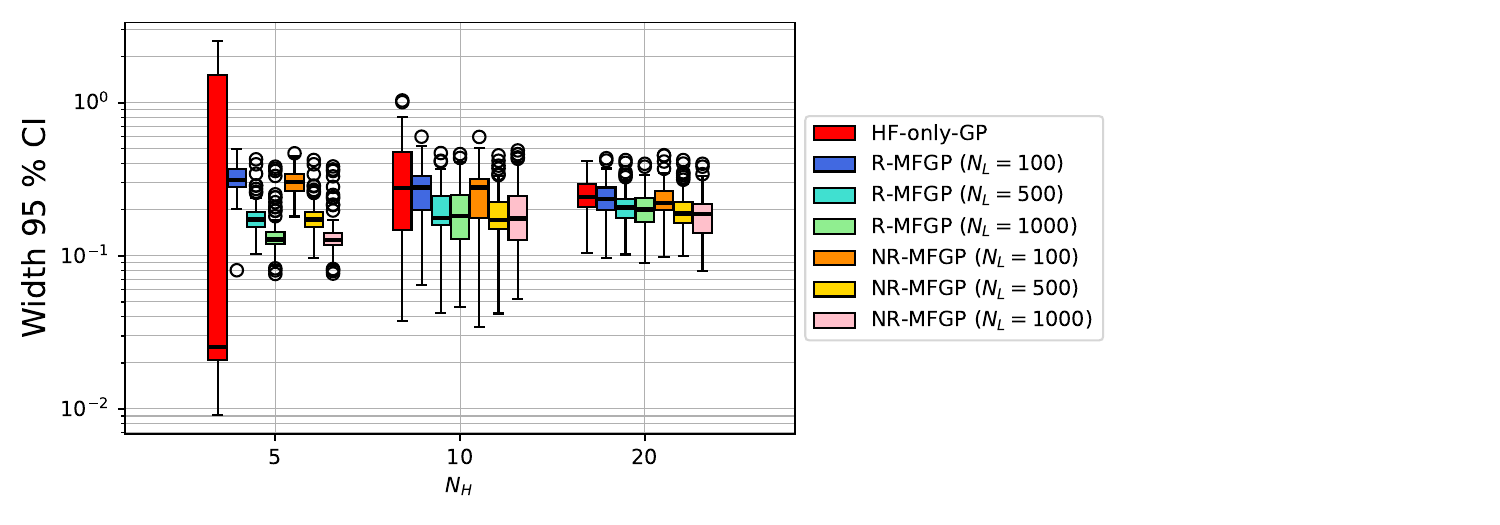}
\caption[]{Mean $95\%$ CI width (log scale) of the HF-only-GP and of the recursive (R-MFGP) and non-recursive (NR-MFGP) formulations. Box-plots are computed over $100$ random designs.}
\label{fig:WCI_sine}
\end{figure}

\begin{figure}[H]
\centering
\includegraphics[height=6.5cm]{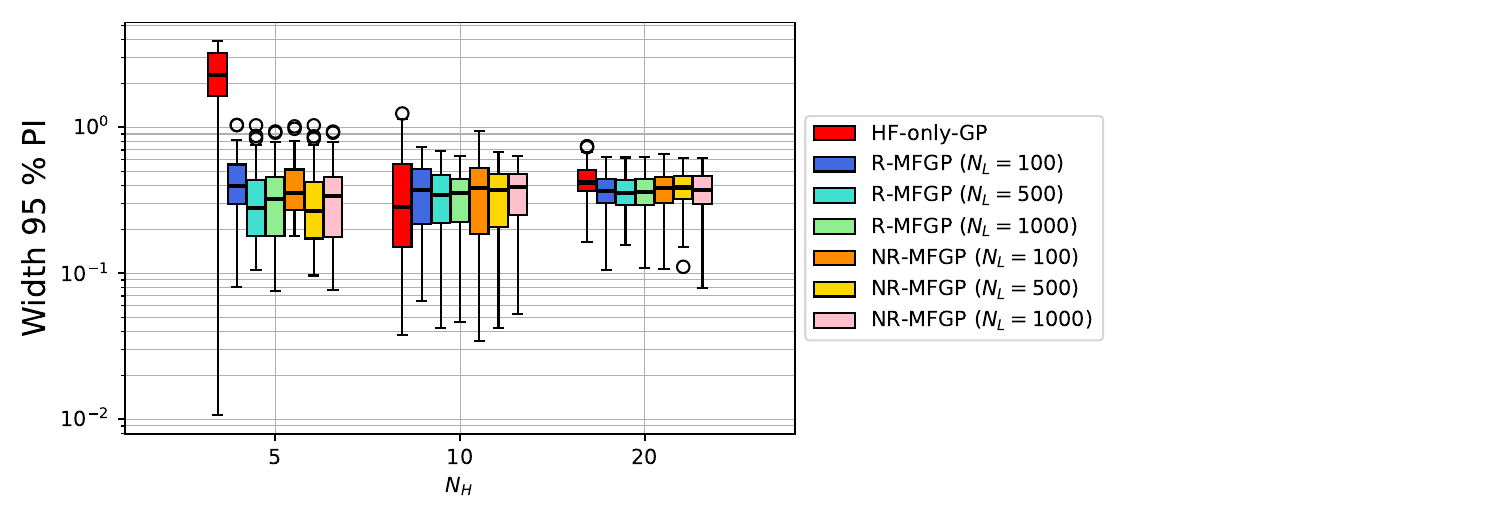}
\caption[]{Mean $95\%$ PI width (log scale) of the HF-only-GP and of the recursive (R-MFGP) and non-recursive (NR-MFGP) formulations. Box-plots are computed over $100$ random designs.}
\label{fig:WPI_sine}
\end{figure}

\begin{figure}[H]
\centering
\includegraphics[height=6.5cm]{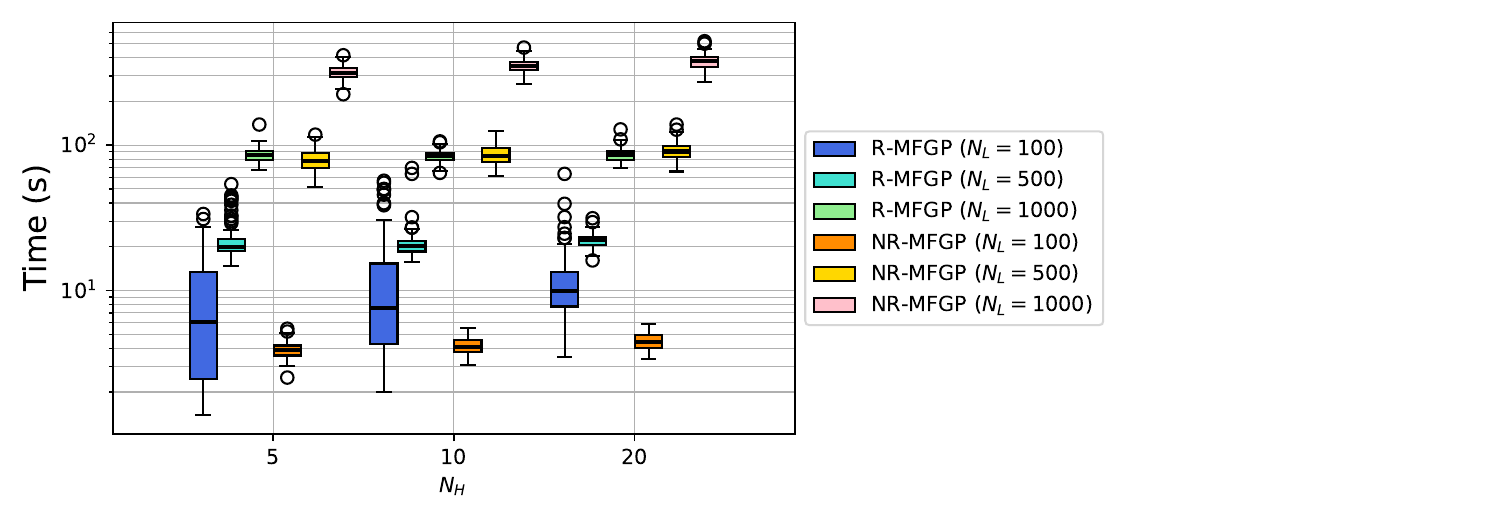}
\caption[]{Wall-clock training time of the recursive (R-MFGP) and non-recursive (NR-MFGP) formulations (in seconds, for $20$ multi-starts). Box-plots are computed over $100$ random designs.}
\label{fig:times_sine}
\end{figure}

The main observations are the following. When comparing the R-MFGP to the NR-MFGP, we notice according to Figures \ref{fig:Errors_sine}, \ref{fig:IAE_CI_sine}, \ref{fig:IAE_PI_sine}, \ref{fig:WCI_sine} and \ref{fig:WPI_sine} that their performances are extremely similar for every performance metric in every configuration. Figure \ref{fig:times_sine} shows that the R-MFGP is notably faster than the NR-MFGP for large LF datasets. In particular, we observe that for $N_L=500$ and $N_L=1000$ the R-MFGP is about $3.5$ to $4$ times faster than the NR-MFGP when considering their median training times. However, the NR-MFGP appears to be clearly faster for $N_L=100$. Table \ref{tab:sine_EM_time} details the computational cost of the R-MFGP model. When $N_L$ increases, the fit of the LFGP becomes more expensive and its share of the total training time also increases. It accounts for more than 95\% of the total time for $N_L = 1000$, whereas for $N_L = 100$ it represents at most 11\%. Conversely, the number of EM iterations, and hence the time spent on the HF parameters, decreases with $N_L$. This is consistent with the theory of the EM algorithm, whose rate of convergence is governed by the fraction of missing information (\cite{Dempster1977EM}). The missing data is the random vector $\widetilde{\mathbf{Y}}\!_{L}(X_{tr}^H)$, whose element-wise variance decreases as $N_L$ increases, or $\sigma^2_{\varepsilon,L}$ decreases for instance. The advantage of the R-MFGP over the NR-MFGP grows with $N_L$. Indeed, for $N_L \gg N_H$, the computational cost of the R-MFGP is dominated by the fit of the LFGP, in $\mathcal{O}(N_L^3)$ per likelihood evaluation. Each likelihood evaluation of the NR-MFGP has a similar cost, in $\mathcal{O}((N_L+N_H)^3)$ but it optimizes numerically more parameters than the LFGP (7 against 2 here), which tends to increase both the cost of each gradient evaluation and the number of iterations of the optimizer.

\begin{table}[H]
\centering
\caption{R-MFGP model: computation time of the LFGP fit and EM algorithm, and the number of EM iterations for each configuration $(N_L, N_H)$. Values are given as median [first quartile, third quartile] over the 100 replications. The EM algorithm is stopped when the relative increment of the HF log-likelihood falls below $10^{-10}$, with a maximum of 200 iterations. This maximum is reached in 31\%, 9\% and 4\% of the runs for $N_H = 5$ and $N_L = 100$, $500$ and $1000$ respectively, in 8\% and 2\% of the runs for $N_H = 10$ and $N_L = 100$ and $500$, and never otherwise.}
\label{tab:sine_EM_time}
\begin{tabular}{ccccc}
\toprule
$N_L$ & $N_H$ & LF fit (s)  & EM (s) & EM iterations \\
\midrule
100  & 5  & 0.7  [0.6, 0.7]    & 5.4 [1.9, 12.7] & 64 [24, 200] \\
500  & 5  & 18.2 [16.9, 19.5] & 1.3 [0.7, 4.5]   & 17 [10, 34]  \\
1000 & 5  & 81.9 [77.3, 86.5]   & 0.8 [0.5, 1.7]   & 10 [7, 18]    \\
\midrule
100  & 10 & 0.6 [0.6, 0.7] & 6.9 [3.6, 14.7] & 28 [20, 61]  \\
500  & 10 & 17.7 [16.5, 19.0] & 2.1 [1.5, 3.1]  & 10 [8, 14]   \\
1000 & 10 & 83.2 [77.3, 87.4] & 1.6 [1.1, 2.3]  & 7 [7, 9]     \\
\midrule
100  & 20  & 0.6 [0.6, 0.7]    & 9.4 [7.1, 12.7] & 26 [21, 38]  \\
500  &  20 & 18.0 [16.5, 19.4] & 4.0 [3.1, 4.7]  & 10 [9, 12]   \\
1000 & 20 & 83.5 [76.5, 88.0] & 3.1 [2.5, 3.6]  & 7 [7, 8]       \\
\bottomrule
\end{tabular}
\end{table}

Figure \ref{fig:Errors_sine} shows that when HF data is scarce, the MFGPs yield better $Q^2$ values compared to the HF-only-GP which is expected. For $N_H=5$, the HF-only-GP gives poor $Q^2$ values and is very unstable overall as shown by Figure \ref{fig:WCI_sine}. For a higher number of HF training points, namely, $N_H=20$, the difference of performance between the HF-only-GP and MFGPs is rather negligible in terms of $Q^2$ and $\text{IAE}_{\text{CI}}$, however, the HF-only-GP tends to be slightly better in terms of $\text{IAE}_{\text{PI}}$.

\subsection{Analytical 4D case: the Park function}
We consider the Park function with:
\[ 
y_H(x_1, x_2, x_3, x_4) = \displaystyle \frac{1}{2} \! \left(\sqrt{x_1^2+(x_2 + x_3^2)x_4} - x_1 \right) + (x_1+3x_4) \exp(1 + \sin(x_3)) 
\]
being the HF test function and for the LF function, we define:
\[ 
y_L(x_1, x_2, x_3, x_4) = \displaystyle \left(1+\frac{\sin(x_1)}{10} \right)y_H(x_1, x_2, x_3, x_4) - 2x_1+x_2^2 +x_3^2 + 0.5. 
\]
Both functions are defined for $\textbf{x} \in [0,1]^4$ and as for their respective ranges, $y_H(\textbf{x})$ takes values between $0$ and $25.59$, and $y_L(\textbf{x})$ is between $0.5$ and $28.25$. The function $y_H$ is the continuous extension to $x_1=0$ of the function proposed by \cite{Cox2001Parkfunction}. The multi-fidelity version is given by \cite{Xiong2013test_fcts} among other analytical test functions. We present a benchmark obtained with many different  experimental designs where $N_L \in \{150, 300, 600\}$, $N_H \in \{ 20, 40, 60\}$ and $N_{test}=10^4$. All training and test points are sampled with maximin LHS on $[0,1]^4$ and artificial Gaussian noise is generated on both fidelity levels with standard deviations $\sigma_{\varepsilon,L}=2.5$ and $\sigma_{\varepsilon,H}=0.5$. We again compare the MFGP AR(1) models to the HF-only-GP. We compute the metric values for $N_{rep}=100$ different samples of the training points and noise realizations. \\

\begin{figure}[H]
    \centering
    \includegraphics[height=6.5cm]{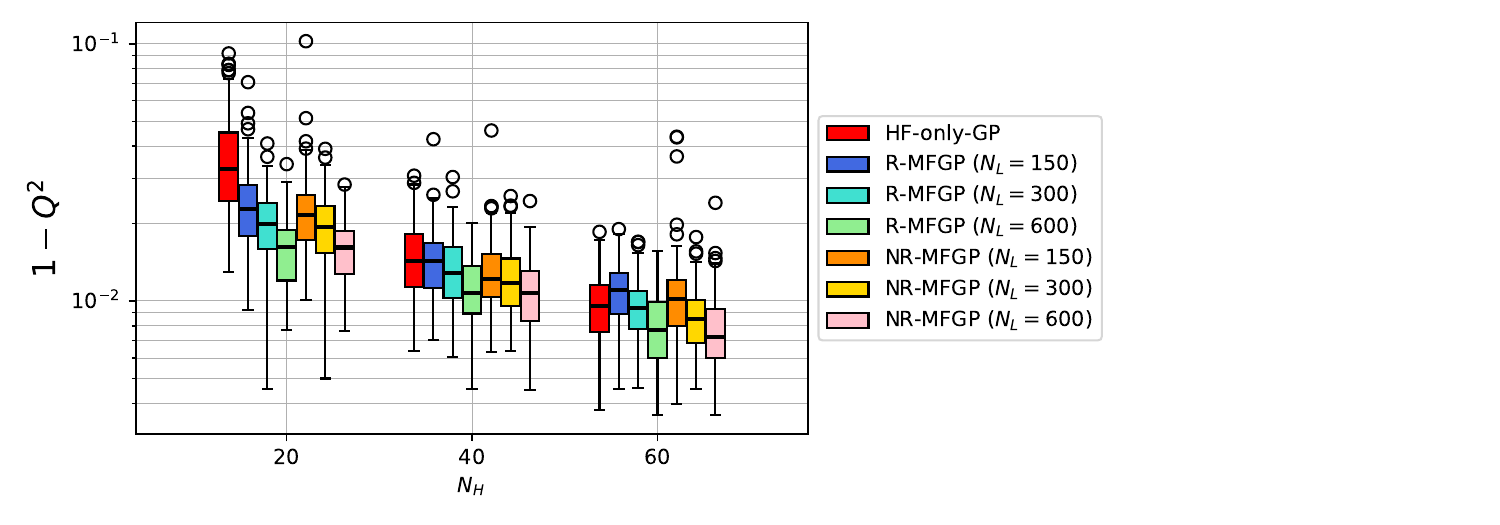}
    \caption[]{Prediction error $1 - Q^2$ (log scale) of the HF-only-GP and of the recursive (R-MFGP) and non-recursive (NR-MFGP) formulations. Box-plots are computed over $100$ random designs.}
    \label{fig:Errors_Park}
\end{figure}

\begin{figure}[H]
    \centering
    \includegraphics[height=6.5cm]{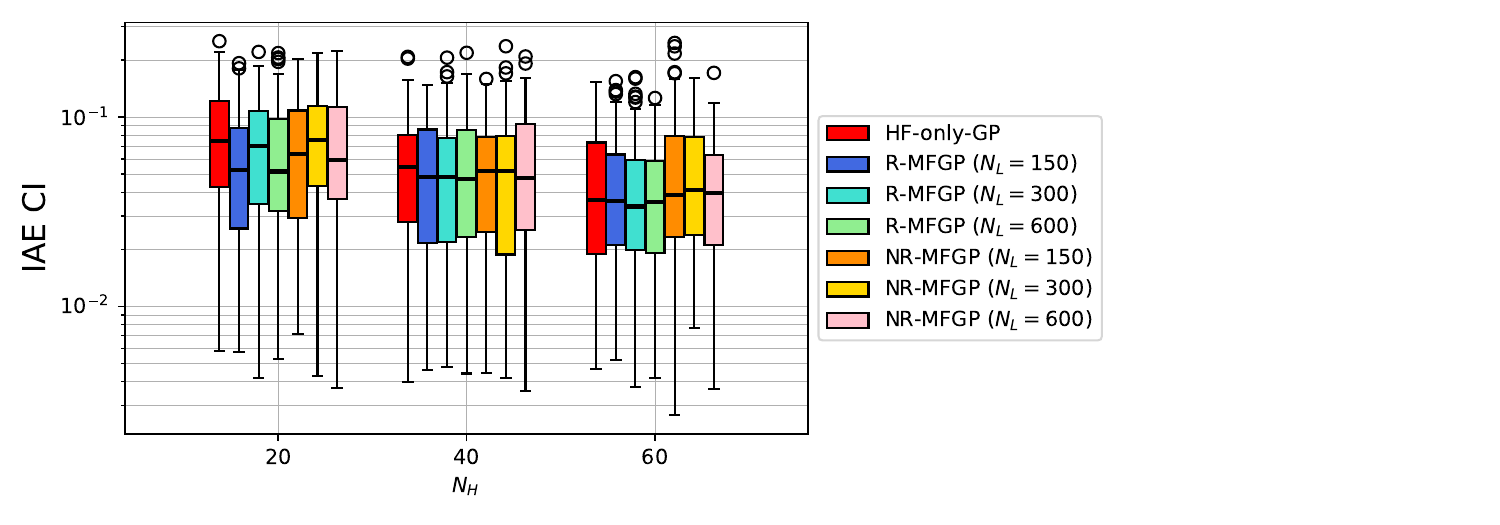}
    \caption[]{$\text{IAE}_{\text{CI}}$ (log scale) of the HF-only-GP and of the recursive (R-MFGP) and non-recursive (NR-MFGP) formulations. Box-plots are computed over $100$ random designs.}
    \label{fig:IAE_CI_Park}
\end{figure}

\begin{figure}[H]
    \centering
    \includegraphics[height=6.5cm]{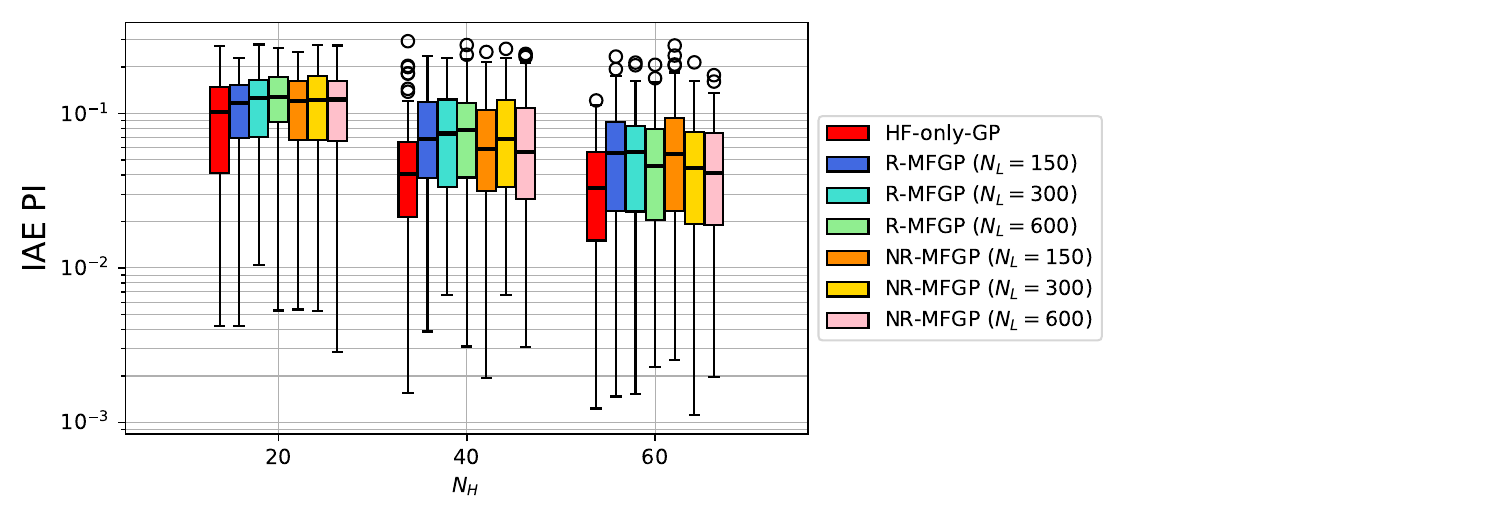}
    \caption[]{$\text{IAE}_{\text{PI}}$ (log scale) of the HF-only-GP and of the recursive (R-MFGP) and non-recursive (NR-MFGP) formulations. Box-plots are computed over $100$ random designs.}
    \label{fig:IAE_PI_Park}
\end{figure}

\begin{figure}[H]
    \centering
    \includegraphics[height=6.5cm]{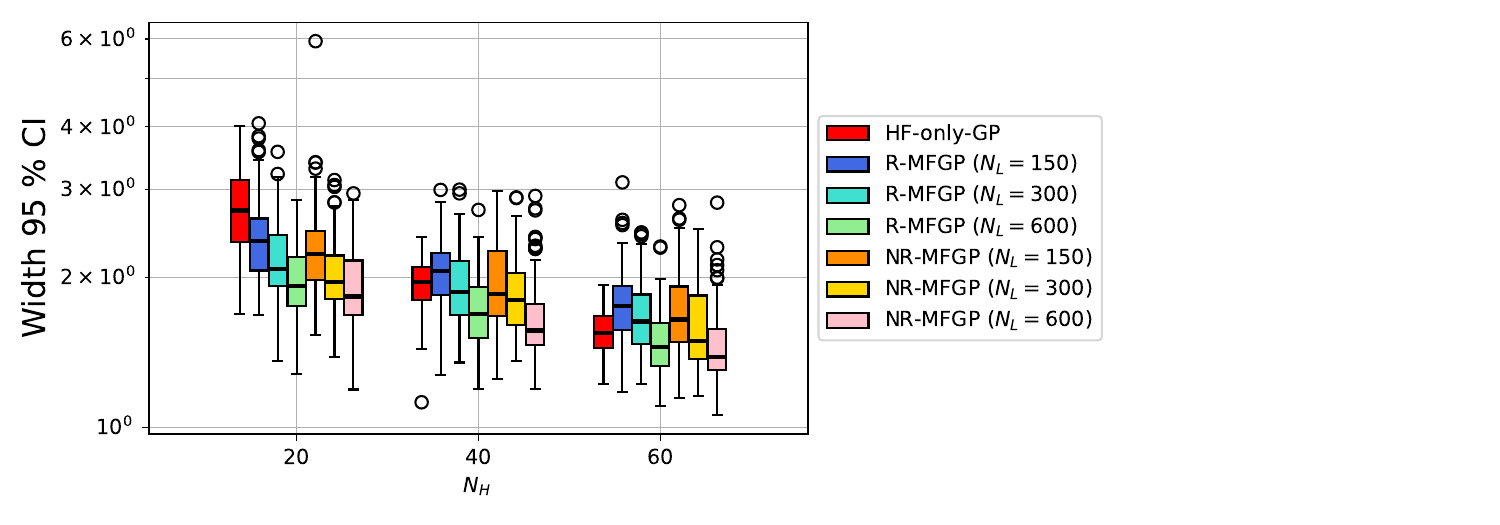}
    \caption[]{Mean $95\%$ CI width (log scale) of the HF-only-GP and of the recursive (R-MFGP) and non-recursive (NR-MFGP) formulations. Box-plots are computed over $100$ random designs.}
    \label{fig:WCI_Park}
\end{figure}

\begin{figure}[H]
    \centering
    \includegraphics[height=6.5cm]{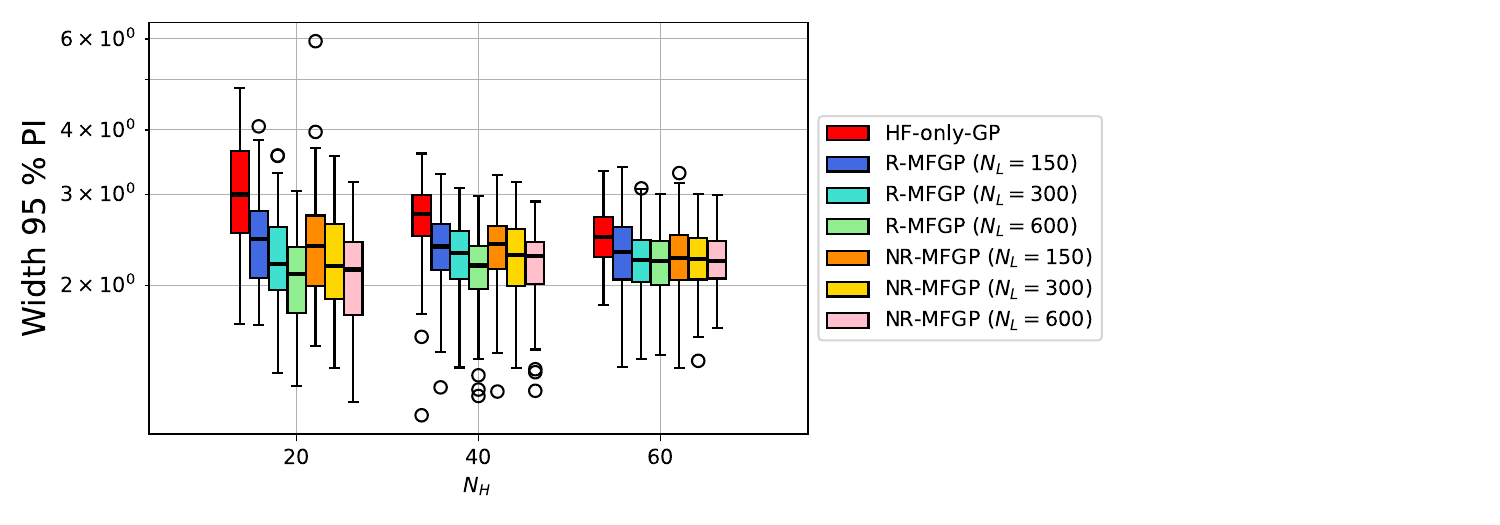}
    \caption[]{Mean $95\%$ PI width (log scale) of the HF-only-GP and of the recursive (R-MFGP) and non-recursive (NR-MFGP) formulations. Box-plots are computed over $100$ random designs.}
    \label{fig:WPI_Park}
\end{figure}

\begin{figure}[H]
    \centering
    \includegraphics[height=6.5cm]{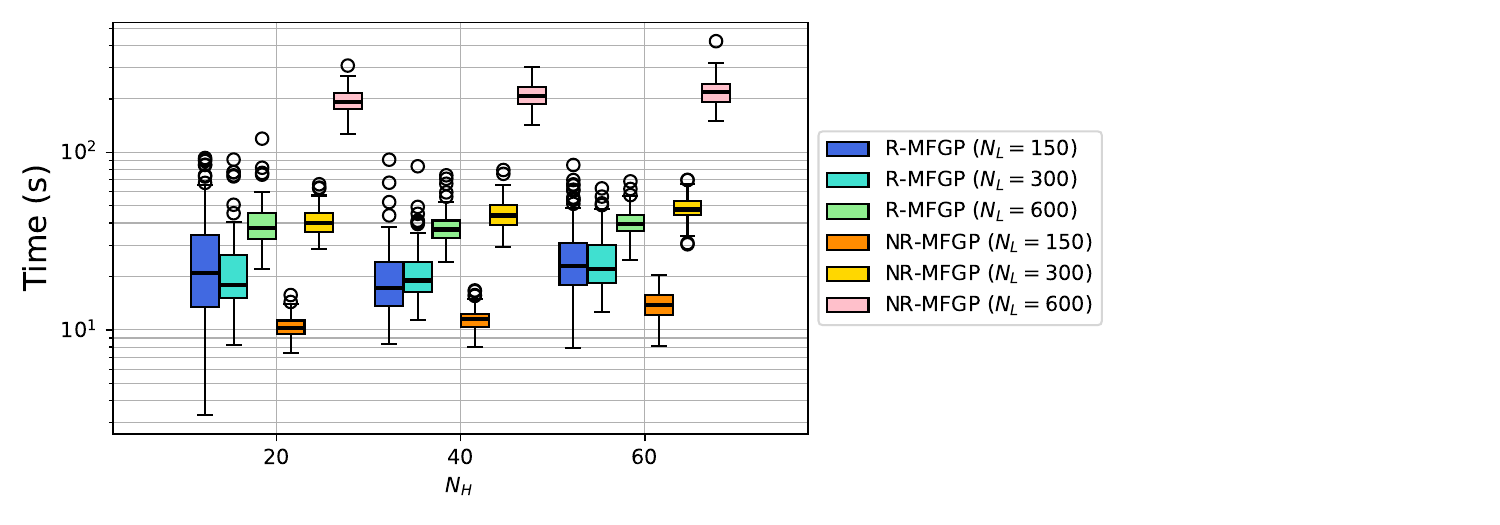}
    \caption[]{Wall-clock training time of the recursive (R-MFGP) and non-recursive (NR-MFGP) formulations (in seconds, for $20$ multi-starts). Box-plots are computed over $100$ random designs.}
    \label{fig:times_Park}
\end{figure}

The main results are as follows. The two MFGP formulations yield almost identical results for every performance metric as shown by Figures \ref{fig:Errors_Park}, \ref{fig:IAE_CI_Park}, \ref{fig:IAE_PI_Park}, \ref{fig:WCI_Park} and \ref{fig:WPI_Park}. As in the previous test case, the main difference between the two models lies in their training times. According to Figure \ref{fig:times_Park}, the R-MFGP is about $5$ times faster for $N_L = 600$, when comparing their median training times, whereas the NR-MFGP is clearly faster for $N_L = 150$. Table \ref{tab:park_EM_time} shows the same behavior as in the previous test case: when $N_L$ increases, the number of EM iterations decreases, while the share of the LFGP fit in the total training time increases, from less than 10\% for $N_L = 150$ to about 75\% for $N_L = 600$. The same explanation also holds here, as the NR-MFGP optimizes numerically 13 parameters against 5 for the LFGP.  \\

\begin{table}[h]
\centering
\caption{R-MFGP model: computation time of the LFGP fit and EM algorithm, and the number of EM iterations for each configuration $(N_L, N_H)$. Values are given as median [first quartile, third quartile] over the 100 replications. The EM algorithm is stopped when the relative increment of the HF log-likelihood falls below $10^{-10}$, with a maximum of 200 iterations. This limit is reached in 7\%, 2\% and 1\% of the runs for $(N_L, N_H) = (150, 20)$, $(300, 20)$ and $(150, 60)$ respectively, and never otherwise.}
\label{tab:park_EM_time}
\begin{tabular}{ccccc}
\toprule
$N_L$ & $N_H$ & LF fit (s) & EM (s) & EM iterations \\
\midrule
150   & 20 &   1.4 [1.3, 1.7] & 18.8 [12.1, 32.6] & 54 [34, 85]  \\
300  &   20 & 6.9 [6.1, 7.6]  & 11.8 [7.4, 19.7]  & 33 [24, 53] \\
600 & 20 & 28.8 [25.3, 32.3] & 8.5  [5.0, 12.8]  & 22 [18, 32] \\
\midrule
150 & 40 & 1.5 [1.3, 1.6]    & 15.7  [11.8, 22.4] & 43 [31, 52]  \\
300 & 40 &  6.6 [5.8, 7.6]    & 12.6 [10.1, 17.4] & 31 [26, 40] \\
600 & 40 & 28.4 [25.4, 31.7] & 7.7 [6.0, 10.7]   & 19 [16, 24] \\
\midrule
150 & 60 &  1.5 [1.4, 1.7]  & 21.5 [16.2, 29.4] & 42 [34, 61] \\
300  & 60 & 6.6 [5.7, 7.6]    & 15.5 [11.7 , 23.0] & 30 [24, 44] \\
600 & 60 & 28.9 [25.8, 33.4] & 9.7 [7.3, 13.2]  & 19 [15, 24]   \\
\bottomrule
\end{tabular}
\end{table}

As shown in Figure \ref{fig:Errors_Park}, the MFGP models outperform the HF-only-GP in terms of $1-Q^2$ values when $N_H$ is low, but this difference in performance becomes negligible when $N_H$ takes higher values. Increasing $N_L$ from $150$ to $600$ offers a slight improvement, but it is not as impactful as increasing $N_H$. Regarding uncertainty quantification (UQ), Figure \ref{fig:IAE_CI_Park} shows that in terms of $\text{IAE}_{\text{CI}}$, the three models give similar results, but the HF-only-GP clearly outperforms the MFGP models on the $\text{IAE}_{\text{PI}}$ metric for $N_H \geq 40$, as shown in Figure \ref{fig:IAE_PI_Park}. Increasing $N_L$ does not seem to help the MF model in that regard. This suggests that if the computational budget is high enough, i.e. the user has a lot of available HF observations, it would be preferable in terms of reliability to use a simple GP model, but only if it yields an error value comparable to the one of the MFGP. The added value of the MF models is especially apparent when few HF samples are given, or in other words, when the budget is limited. \\

Figures \ref{fig:WCI_Park} and \ref{fig:WPI_Park} also provide results on the mean CI and PI widths for $\alpha=0.95$. The mean CI widths tend to decrease and the mean PI widths seem to be more stable when $N_H$ increases for all three models. This indicates that the share of the estimated noise in the predictive variance grows with $N_H$. For the HF-only-GP, we also observe an improvement of the $\text{IAE}_{\text{CI}}$ and $\text{IAE}_{\text{PI}}$. For higher values of $N_H$, the variability over the mean CI and PI widths across designs is clearly smaller. We propose the following explanation: for small $N_H$, the variability of the HF data is mostly attributed to the GP kernel rather than to the noise, and that this split becomes more accurate as $N_H$ increases. The same reasoning could be applied to the MF models as their IAE values also decrease, but slower. Thus, the UQ provided by the MFGPs is improved but not as rapidly as for the HF-only-GP.

\subsection{Real-world case: sea surface temperature dataset}

We now apply both multi-fidelity GP models to a real dataset of sea surface temperatures (SST) measurements in the Massachusetts and Cape Cod bays, which is detailed by \cite{Babaee2020CapeCod}. We are interested in providing spatiotemporal predictions, thus, the inputs are triplets (longitude, latitude, time) and the output is the SST. The HF data consists of Massachusetts Water Resources Authority (MWRA) measurements at $14$ different stations in the Massachusetts and Cape Cod Bays. The temperature is measured one meter below the surface of the seawater, which is taken as a proxy for the SST, and the measurements are taken every month, except for the winter months, for a time span of two years (2015-2016). The LF data consists of images of 4 km $\times$ 4 km resolution from the Moderate-resolution Imaging Spectroradiometer (MODIS) aboard NASA's Terra satellite, further details are provided by \cite{Babaee2020CapeCod}. Among the $14$ MWRA stations, $11$ are used for training and $3$ are used for testing, which results in $N_H=195$ HF training points. \cite{Babaee2020CapeCod} consider monthly satellite images, hence, $N_L=2526$ LF data points are present in the training set. In terms of numerical optimization, the recursive approach solves a 4D problem at the LF level and a sequence of 4D problems at the HF level, instead of a global 11D problem for the non-recursive formulation. Regarding the NR-MFGP, the difference with \cite{Babaee2020CapeCod} is that the parameters $\beta_L$ and $\beta_H$ are estimated from the data and not fixed at zero. We consider $10$ multi-starts for each optimization. To assess the computational cost and stability, the training of each formulation is repeated $5$ times. Training times are stable across runs, with a coefficient of variation of $1\%$ for both formulations. The median wall-clock training time is $13.5$ minutes for the recursive formulation against $68$ minutes for the non-recursive one, i.e. a reduction by a factor of about $5$. \\

\begin{figure}[H]
\centering
\hspace*{-2cm}
\includegraphics[height=11.2cm]{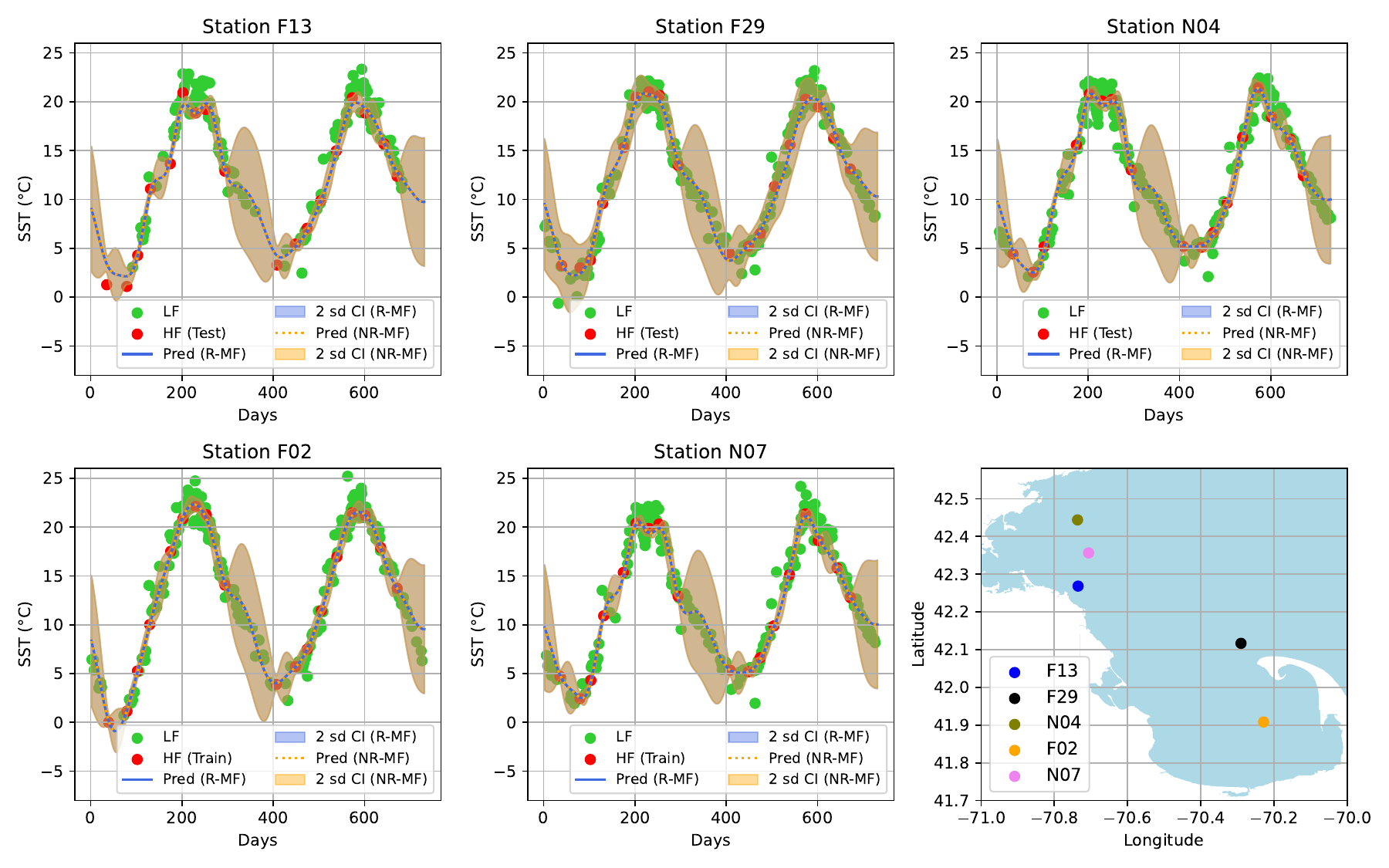}
\caption[]{Multi-fidelity predictions and credible intervals for both models (R-MFGP in blue, NR-MFGP in orange) at each day of the 2015-2016 period for the five stations: F13, F29, N04, F02, N07. The first row corresponds to the test stations, i.e. their data is not in the HF training set. The first two plots of the second row correspond to stations where the HF points are part of the HF training set. The bottom-right plot shows the locations of the stations, the shapes of the bays are obtained using the data from \cite{capecod_dataset}.}
    \label{fig:codmass_time}
\end{figure}

We now present two kinds of results: SST predictions with respect to time at a few specific locations, and SST predictions with respect to longitude and latitude at a particular day. For the predictions with respect to time, $5$ locations are considered, which correspond to the stations called F13, F29, N04, F02 and N07. The data of the first three stations is not part of the HF training set, hence, it can be seen as a test set. This is not the case for the data points obtained at the last two stations, which are present among the HF training points. Table \ref{tab:capecod-stations} shows that the two MFGP formulations yield near-identical results for the test stations. The MFGPs achieve slightly higher $Q^2$ values than the HF-only-GP, whereas the latter gives a lower $\text{IAE}_{\text{PI}}$ at two of the three stations. These differences should however be interpreted with caution, given the small number of measurements per station. Figure \ref{fig:codmass_time} shows that the MF predictions closely follow the trend of the LF and HF data points for all $5$ stations. The predictions and credible intervals are almost identical for both MFGP formulations. Overall, the predictions are similar to those given in \citet[p.~11]{Babaee2020CapeCod}, except for the first days, which can be explained by the fact that no LF training data is available before the 120th day. In particular, not all the LF points shown in Figure \ref{fig:codmass_time} are part of the training set, due to the fact that only monthly satellite images are considered for training. The LFGP tends to yield near-constant predictions equal to its estimated prior mean value for the first days, which is learned from the data and not fixed in our application. The other difference is that we observe larger credible intervals between day $300$ and day $400$, as it corresponds to a time period during which no HF training point is available. \\

\begin{figure}[H]
    \centering
    \hspace*{-0.95cm}
\includegraphics[height=13.cm]{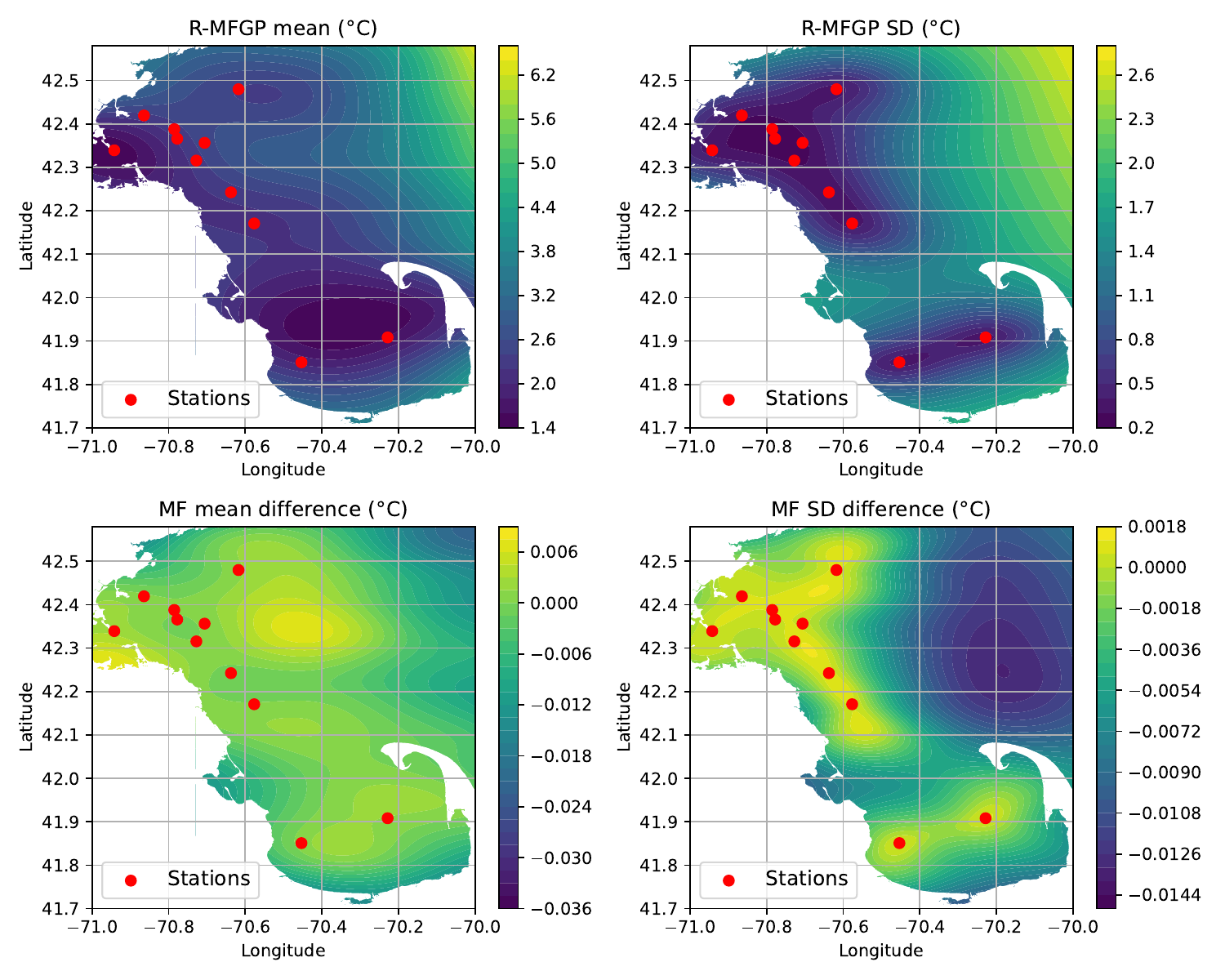}
    \caption[]{Top row: recursive MFGP mean predictions (left) and standard deviations (right) computed on a regular $100 \times 100$ grid of $[-71,-70] \times [41.7, 42.6]$ for March 22, 2015. Bottom row: Difference between the recursive and non-recursive predictions for the mean (left) and standard deviation (right) on the same grid. The red dots correspond to the locations of the MWRA stations whose data is used in the HF training set. The shapes of the bays are obtained using the data from \cite{capecod_dataset}.}
    \label{fig:codmass_space}
\end{figure}

Regarding the predictions with respect to spatial coordinates, we fix the time component at a given day, in this case: March 22, 2015. We compute the MF posterior means and standard deviations on a $100 \times 100$ spatial grid. Figure \ref{fig:codmass_space} shows that the R-MFGP and NR-MFGP yield close results in terms of the posterior mean and standard deviation. Our predictions are again similar to those of \citet[p.~13]{Babaee2020CapeCod} for the same day, in terms of both MF posterior mean and standard deviation values. For the MF mean, the same patterns can be observed, notably that colder temperatures are predicted around the coordinates $(-70.9, 42.35)$  and also near the center of Cape Cod bay, in the region $[-70.4, -70.2] \times [41.9, 42.0]$. As expected, the MF standard deviations are the smallest around the MWRA stations. Although we present here only the result for March 22, we did the same computations for other days: January 1st, January 31st, March 2nd, April 1st and May 1st. For all tested days, the posterior mean and standard deviations patterns are analogous to the results of \cite{Babaee2020CapeCod}, however, for January 1st especially, but also January 31st, the temperature ranges are different, this is also due to the fact that the mean predictions are different compared with \citet[p.~11]{Babaee2020CapeCod} at the boundary of the time domain. 

\begin{table}[H]
\centering
\caption{Prediction metrics for
HF-only-GP (HF), R-MFGP (R) and NR-MFGP (NR) at the three test stations F13, F29 and N04. Metrics are computed using the measurements of each station as the target values (18, 18 and 17 values are available for F13, F29 and N04 respectively).}
\label{tab:capecod-stations}
\small
\setlength{\tabcolsep}{4pt}
\begin{tabular}{c ccc ccc ccc}
\toprule
 & \multicolumn{3}{c}{$Q^2$}
 & \multicolumn{3}{c}{$\text{IAE}_{\text{PI}}$}
 & \multicolumn{3}{c}{$\text{MPIW}_{95\%}$} \\
\cmidrule(lr){2-4} \cmidrule(lr){5-7} \cmidrule(lr){8-10}
Station & HF & R & NR & HF & R & NR & HF & R & NR \\
\midrule
F13 & 0.982 & 0.986 & 0.987 & 0.084 & 0.063 & 0.064 & 1.898 & 1.975 & 1.971 \\
F29 & 0.980 & 0.992 & 0.992 & 0.081 & 0.122 & 0.123 & 3.556 & 4.076 & 4.087 \\
N04 & 0.998 & 0.999 & 0.999 & 0.158 & 0.189 & 0.188 & 1.869 & 1.901 & 1.896 \\
\bottomrule
\end{tabular}
\end{table}

\section{Conclusion}

We investigated the recursive formulation of the multi-fidelity auto-regressive GP model in the general setting of noisy outputs and non-nested experimental designs. We developed a decoupled optimization strategy that exploits the EM algorithm as in \cite{zertuche_thesis} but in the framework of noisy observations. This strategy addresses the limitation reported in \cite{Babaee2020CapeCod} by substantially reducing the dimension of each optimization problem. We derived a closed-form update formula when the scaling factor $\rho$ is a parametric linear predictor function. The proposed approach was compared to the fully coupled likelihood maximization of the non-recursive formulation on several applications cases. We observed in our test cases that the decoupled strategy with the EM algorithm achieved similar predictions to those of the non-recursive formulation, with faster training times when large amounts of LF data are available. For small LF datasets, however, the non-recursive formulation was faster. In practice, the former situation is the most common in multi-fidelity applications, since LF data is cheap and typically available in large numbers. Fast estimation strategies are especially useful when multi-start optimizations are needed in order to improve the reliability of the result. In the context of noisy observations, a simple GP model can be preferable to a MFGP in some configurations, notably in terms of uncertainty estimation, but only when a high number of HF data points are available. When the HF computational budget is limited, the MFGP models yield the smallest error values. A natural extension would be the study of more complex noise models with input-dependent noise variances (\cite{TAO2025Het_MFGP}). 

\section*{Acknowledgments}

This research was supported by the CEA (French Alternative Energies and Atomic Energy Commission) and the SEISM Institute (\url{https://www.institut-seism.fr/en/}).

\section*{Data availability}

The code and data needed to reproduce the results are available in the Github repository: \url{https://github.com/NilsBaillie/MFGP-noisy-nonnested}.

\section*{Declaration of generative AI and AI-assisted technologies in the manuscript preparation process}

During the preparation of this work the authors used Claude to improve readability and to assist with software development. After using Claude, the authors reviewed and edited the content as needed and take full responsibility for the content of the published article.

\bibliographystyle{plainnat} 
\bibliography{biblio}

\appendix

\section{Implementation remarks}\label{app:implementation}

All numerical experiments were run on a Dell laptop with an Intel Core Ultra 9 185H CPU (16 cores, 22 threads) with 32 GB of RAM using Python 3.13.2, Numpy (2.1.3) and Scipy (1.15.2). Linear algebra operations rely on the default multi-threaded OpenBLAS backend (22 threads). All methods were run under identical conditions.  \\

For the EM algorithm, we set the maximum number of iterations to $200$. The EM algorithm can stop early if the relative increment $|\ell_t - \ell_{t-1}|/ \max(|\ell_{t-1}|,|\ell_{t}|,1)$ is less than $10^{-10}$, where $\ell_t$ is the HF log-likelihood value at iteration $t$. This stopping criterion has the same form as the one used by default in Scipy. \\

In practice, slightly different parametrizations of the GP models are used for the numerical optimization step. For the LF part in the recursive approach, we consider $(\boldsymbol \theta_L, \log\eta_L)$. Likewise, in the training of the HF-only-GP and in every M-step for the recursive model, we optimize with respect to $(\boldsymbol \theta_H, \log\eta_H)$.  The fully coupled likelihood in the non-recursive formulation is optimized with respect to $\beta_{\rho}$, $\log \sigma^2_L, \log \sigma^2_H$, $\boldsymbol\theta_L$, $\boldsymbol\theta_H$, $\log \sigma^2_{\varepsilon,L}$ and $\log \sigma^2_{\varepsilon,H}$. We observed that using the log rather than the original parameter yields much more robust estimates. \\ 

In order to avoid unstable estimations, some bounds are imposed on the parameters concerned by numerical optimization. The bounds are used as box-constraints in the L-BFGS-B algorithm. The same type of bounds are considered for the LF and HF versions of a given parameter. For the variance-type parameters,  $\log \eta \in [-40, 10]$, $\log \sigma^2 \in [-60, 10]$, $\log \sigma^2_{\varepsilon} \in [-40, 10]$ act as numerical safeguards. Similarly, $\beta_{\rho} \in [-10, 10]$ covers a wide range of LF-HF scaling factors in particular for normalized outputs. The correlation lengths are bounded by the range supported by the data: $\theta_d \in \left[\theta_{\inf}^{(d)},  \theta_{\sup}^{(d)} \right]$ where:
\[  
\theta_{\inf}^{(d)}=\min_{x_i^{(d)} \neq x_j^{(d)}}  \left|x_i^{(d)} - x_j^{(d)} \right|, \quad \theta_{\sup}^{(d)}= \left(\max_i x_i^{(d)} - \min_i x_i^{(d)} \right)
\]
with $x_i^{(d)}$ being the $d$-th coordinate of the $i$-th element in the input training set. The bounds are also used when sampling the different initial points for the multi-start optimization with maximin LHS. For the 1D and 4D test function cases, we use 20 multi-start initial points. This number is reduced to 10 points for the real-world case to keep a reasonable training time. \\

The EM algorithm is dependent on the initialization in practice. We use the values $\beta_{\rho}=1$, $\sigma^2_H=1$, $\beta_H=0$, $\eta_H=1$ and $\theta_{H,d}=\theta_{\sup}^{(d)}/2$ (or the midpoint when the function is defined on a known bounded interval). We consider that these values are a natural first guess, even more so if the user normalizes their data before applying the model. This particular point is present among the other initial points in the multi-start optimization for the HF-only-GP and non-recursive MFGP models.

\section{Detailed derivation of the E-step}\label{app:EM_details}

The function $Q_{alt}$ is defined from the distribution of $(\widetilde{\mathbf{Y}}\!_{L}(X_{tr}^H) \, | \, \textbf{Z}_H(X_{tr}^H))$, which can be obtained from the joint distribution and using the Gaussian conditioning theorem. We also suppose that $\widetilde{Y}_L$ and $\varepsilon_H$ are independent. At $\boldsymbol\xi_H = \boldsymbol\xi_H^{(t)}$, we have:
$$\begin{pmatrix}
\widetilde{\mathbf{Y}}\!_{L}(X_{tr}^H) \\
\textbf{Z}_H(X_{tr}^H) \\
\end{pmatrix}  \sim \mathcal{N}\!\left( \begin{pmatrix}
\textbf{m}_{Y_L}(X_{tr}^H)  \\
\mathrm{D}_\rho \textbf{m}_{Y_L}(X_{tr}^H) +  F_H\boldsymbol\beta_H^{(t)} \\
\end{pmatrix}, \begin{pmatrix}
V_{Y_L}(X^H_{tr},X^H_{tr}) & \Sigma_{\textbf{YZ}}^{(t)}  \\
(\Sigma_{\textbf{YZ}}^{(t)})^\top &  \Sigma_{\textbf{ZZ}}^{(t)}
\end{pmatrix} \right)\!,  $$
then, the Gaussian conditioning theorem yields:
$$ \left( \widetilde{\mathbf{Y}}\!_{L}(X_{tr}^H)
\, | \, \textbf{Z}_H(X_{tr}^H)=\textbf{z}_{tr}^H; \boldsymbol\xi_H^{(t)} \right) \sim \mathcal{N}\!\left(\boldsymbol\mu_{\textbf{Y}|\textbf{Z}}^{(t)}, \Sigma_{\textbf{Y}|\textbf{Z}}^{(t)} \right)\!, $$
the expressions for $\Sigma_{\textbf{YZ}}^{(t)}$, $\Sigma_{\textbf{ZZ}}^{(t)}$, $\boldsymbol\mu_{\textbf{Y}|\textbf{Z}}^{(t)}$ and $\Sigma_{\textbf{Y}|\textbf{Z}}^{(t)}$ are given by (\ref{eq:Many_Matrices}). Thus, the objective function to be maximized at the M-step is of the form: 
\begin{align*}
Q_{alt}(\boldsymbol\xi_H;\boldsymbol\xi_H^{(t)}) &=  -\frac{N_H}{2}\log(\sigma_H^2) - \frac{1}{2}\log \det \!\left(R_H(\boldsymbol\theta_H) + \eta_HI_{N_H}\right)-\frac{N_H}{2}\log(2\pi) \\
&- \frac{1}{2\sigma_H^2}\mathbb{E}_{( \widetilde{\mathbf{Y}}\!_{L}(X_{tr}^H)
 \, | \,\textbf{Z}_H(X_{tr}^H) =\textbf{z}_{tr}^H; \boldsymbol\xi_H^{(t)})}\left[\widetilde{\mathcal{S}}(\widetilde{\mathbf{Y}}\!_{L}(X_{tr}^H), \textbf{z}_{tr}^H, \bm{\xi}_H) \right]\!, 
\end{align*}
with: $\widetilde{\mathcal{S}}(\widetilde{\mathbf{Y}}\!_{L}(X_{tr}^H), \textbf{z}_{tr}^H, \bm{\xi}_H) = \mathcal{S}\!\left(\textbf{z}_{tr}^H - F_H\boldsymbol\beta_H - \mathrm{D}_\rho \widetilde{\mathbf{Y}}\!_{L}(X_{tr}^H); R_H(\boldsymbol\theta_H) + \eta_HI_{N_H}\right)$. The computation of the expectation term follows from the following lemma:
\begin{lemma}
\label{lem:quadratic_expectation}
    Let $\mathbf{Y} \in \mathbb{R}^{N \times 1}$ be a random vector of mean $\boldsymbol\mu$ and covariance matrix $\Sigma$. Suppose we have a fixed matrix $A \in \mathbb{R}^{N \times N}$. We have the formula: 
    \[  \mathbb{E}[\mathbf{Y}^\top A \mathbf{Y}] = \boldsymbol\mu^\top A \boldsymbol\mu + \Tr(A\Sigma).  \]
\end{lemma}
We have by the previous lemma:
\begin{align*}
\mathbb{E}_{( \widetilde{\mathbf{Y}}\!_{L}(X_{tr}^H)
 \, | \,\textbf{Z}_H(X_{tr}^H) =\textbf{z}_{tr}^H; \boldsymbol\xi_H^{(t)})} &\left[\mathcal{S}\!\left(\textbf{z}_{tr}^H - F_H\boldsymbol\beta_H - \mathrm{D}_\rho  \widetilde{\mathbf{Y}}\!_{L}(X_{tr}^H); R_H(\boldsymbol\theta_H) + \eta_HI_{N_H}\right) \right] \\ &=  \Tr \!\left(\mathrm{D}_\rho (R_H(\boldsymbol\theta_H) + \eta_H I_{N_H})^{-1}\mathrm{D}_\rho\Sigma_{\textbf{Y}|\textbf{Z}}^{(t)} \right) \\
&+ \mathcal{S}\!\left(\textbf{z}_{tr}^H - F_H\boldsymbol\beta_H - \mathrm{D}_\rho \boldsymbol\mu_{\textbf{Y}|\textbf{Z}}^{(t)}; R_H(\boldsymbol\theta_H) + \eta_HI_{N_H}\right)\!.
\end{align*}
Since $\mathrm{D}_\rho = \Diag(G_L\boldsymbol\beta_{\rho})$, we can write differently the trace term using a result from \citet[p.~479]{Horn2012Matrix_Analysis} (lemma 7.5.2):
\begin{lemma}
\label{lem:hadamard_prod}
Let $\mathbf{x}, \mathbf{y} \in \mathbb{R}^{N \times 1}$ be column vectors and $A, B \in \mathbb{R}^{N \times N}$ some matrices. We have the formula: 
\[   \Tr\!\left(\Diag(\mathbf{x})^\top A \Diag(\mathbf{y}) B^\top  \right) = \mathbf{x}^\top (A\odot B)\mathbf{y}.\]
\end{lemma}
Thus, with lemma \ref{lem:hadamard_prod} and the different block matrices defined in (\ref{eq:EM_blockmatrices}), we retrieve the given expression for $Q_{alt}$.
\\

We now present the gradient formula of $\widetilde{Q}_{alt}$. For more compact notations, we let $\widetilde{R}_H=R_H(\boldsymbol\theta_H)+\eta_H I_{N_H}$. We have that:
\begin{align*}
\frac{\partial \widetilde{Q}_{alt}}{\partial \omega_H}\!\left((\boldsymbol\theta_H,\eta_H); \bm{\xi}_H^{(t)}\right) &= \frac{1}{2}\Tr \!\left( \!\left( \kappa \kappa^\top  - \widetilde{R}_H^{-1}\right) \frac{\partial \widetilde{R}_H}{\partial \omega_H}\right) \\
&+ \frac{1}{2\hat{\sigma}^{2 (t+1)}_H}\hat{\rho}^{(t+1)}(X_{tr}^H)^\top \!\left((\widetilde{R}_H^{-1} \frac{\partial \widetilde{R}_H}{\partial \omega_H}\widetilde{R}_H^{-1}) \odot \Sigma_{\textbf{Y}|\textbf{Z}}^{(t)} \right)\hat{\rho}^{(t+1)}(X_{tr}^H)
\end{align*}
where $\kappa = \widetilde{R}_H^{-1} \! \left( \textbf{z}_{tr}^H -H_H^{(t)}\hat{\boldsymbol\beta}_{\rho,H}^{(t+1)} \right) / \sqrt{\hat{\sigma}^{2 (t+1)}_H}$ and $\omega_H \in \{ \theta_H^{(1)},\dots, \theta_H^{(D)}, \eta_H \}$.

\section{Proof of theorem \ref{thm:equiv_rec_nrec}}\label{app:proof_thm}

All parameters being fixed, we drop the dependence on $\boldsymbol\theta_L$ and $\boldsymbol\theta_H$ in the notations. We introduce:
\[
A = \sigma_L^2 R_L + \sigma^2_{\varepsilon,L} I_{N_L} = \sigma_L^2 \! \left(R_L + \eta_L I_{N_L}\right), \qquad
B = \sigma_L^2\, r_L(X^L_{tr},X^H_{tr})\, \mathrm{D}_\rho,
\]
\[
C = \sigma_L^2\, \mathrm{D}_\rho\, r_L(X^H_{tr},X^H_{tr})\, \mathrm{D}_\rho + \sigma_H^2 R_H + \sigma^2_{\varepsilon,H} I_{N_H},
\]
so that $\widetilde{V}_{LH} = \begin{pmatrix} A & B \\ B^\top & C\end{pmatrix}$. The matrix $A$ is SPD by assumption, and $S = C - B^\top A^{-1} B$ is the Schur complement of $A$ in $\widetilde{V}_{LH}$. We can write the following decomposition:
\[
\widetilde{V}_{LH} = L \begin{pmatrix} A & \mathbf{0}_{N_L \times N_H} \\ \mathbf{0}_{N_H \times N_L} & S \end{pmatrix} L^\top,
\qquad
L = \begin{pmatrix} I_{N_L} & \mathbf{0}_{N_L \times N_H} \\ B^\top A^{-1} & I_{N_H} \end{pmatrix}\!.
\]
In particular, $\widetilde{V}_{LH}$ is invertible if and only if $S$ is invertible, and the inverse is given by: 
\[ 
\widetilde{V}_{LH}^{-1} = L^{-\top}  \begin{pmatrix} A^{-1} & \mathbf{0}_{N_L \times N_H} \\ \mathbf{0}_{N_H \times N_L} & S^{-1} \end{pmatrix} L^{-1}, 
\qquad 
L^{-1} = \begin{pmatrix} I_{N_L} & \mathbf{0}_{N_L \times N_H} \\ -B^\top A^{-1} & I_{N_H} \end{pmatrix}\!. 
\]
Hence, for all $\textbf{u} = (\textbf{u}_L^\top,\textbf{u}_H^\top)^\top, \textbf{w} = (\textbf{w}_L^\top,\textbf{w}_H^\top)^\top \in \mathbb{R}^{N_L+N_H}$,
\begin{equation}\label{eq:schur_bilinear}
\textbf{u}^\top \widetilde{V}_{LH}^{-1}\textbf{w}
= \textbf{u}_L^\top A^{-1}\textbf{w}_L
+ \left(\textbf{u}_H - B^\top A^{-1}\textbf{u}_L\right)^\top \! S^{-1} \!\left(\textbf{w}_H - B^\top A^{-1}\textbf{w}_L\right)\!.
\end{equation}

The LF posterior covariance evaluated at the HF input points writes:
\[ 
V_{Y_L}(X^H_{tr},X^H_{tr}) = \sigma_L^2 \!\left(  r_L(X^H_{tr},X^H_{tr}) -  r_L(X^H_{tr},X^L_{tr}) \left(R_L + \eta_L I_{N_L}\right)^{-1} r_L(X^L_{tr},X^H_{tr}) \right)\!,
\]
thus, we retrieve $S = \mathrm{D}_\rho V_{Y_L}(X^H_{tr},X^H_{tr}) \mathrm{D}_\rho + \sigma_H^2 R_H + \sigma^2_{\varepsilon,H} I_{N_H} = K_{AR} + \sigma^2_{\varepsilon,H} I_{N_H}$. The matrix inverted in the recursive formulation is exactly the Schur complement of the LF block in the matrix inverted in the non-recursive formulation. We write the cross-covariance vector as: $\textbf{v}_{LH}(\textbf{x}) = (\textbf{a}(\textbf{x})^\top, \textbf{b}(\textbf{x})^\top)^\top$ with:
\[    
\begin{cases}
    \textbf{a}(\textbf{x}) = \rho(\textbf{x})\,\sigma_L^2\, \textbf{r}_L(\textbf{x}) \in \mathbb{R}^{N_L \times 1} \\
    \textbf{b}(\textbf{x}) = \rho(\textbf{x})\,\sigma_L^2\, \mathrm{D}_\rho\, r_L(X^H_{tr},\textbf{x}) + \sigma_H^2\, \textbf{r}_H(\textbf{x}) \in \mathbb{R}^{N_H \times 1}.
\end{cases}
\]
We have that for all $\textbf{x},\textbf{x}' \in \mathbb{R}^D$:
\[  
\begin{cases}
\textbf{b}(\textbf{x}) - B^\top A^{-1}\textbf{a}(\textbf{x}) = \rho(\textbf{x})\, \mathrm{D}_\rho\, v_{Y_L}(X^H_{tr},\textbf{x}) + \sigma_H^2 \textbf{r}_H(\textbf{x})
= \textbf{k}_{AR}(\textbf{x})  \\
\textbf{a}(\textbf{x})^\top A^{-1}\textbf{a}(\textbf{x}') = \rho(\textbf{x})\rho(\textbf{x}')\sigma_L^2\, \textbf{r}_L(\textbf{x})^\top \!\left(R_L + \eta_L I_{N_L}\right)^{-1}\textbf{r}_L(\textbf{x}').
\end{cases}
\]
Applying \eqref{eq:schur_bilinear} with $\textbf{u} = \textbf{v}_{LH}(\textbf{x})$ and $\textbf{w} = \textbf{v}_{LH}(\textbf{x}')$ yields:
\begin{align*}
    \textbf{v}_{LH}(\textbf{x})^\top \widetilde{V}_{LH}^{-1}\textbf{v}_{LH}(\textbf{x}')
&= \rho(\textbf{x})\rho(\textbf{x}')\,\sigma_L^2\, \textbf{r}_L(\textbf{x})^\top \! \left(R_L + \eta_L I_{N_L}\right)^{-1}\textbf{r}_L(\textbf{x}') \\
&+ \textbf{k}_{AR}(\textbf{x})^\top \!\left(K_{AR} + \sigma^2_{\varepsilon,H} I_{N_H}\right)^{-1}\textbf{k}_{AR}(\textbf{x}').
\end{align*}
Plugging this into the expression of $v_{Y_H^{KO}}$, we obtain the equality of the posterior covariance functions:
\begin{align*}
v_{Y_H^{KO}}(\textbf{x},\textbf{x}')
&= \rho(\textbf{x})\rho(\textbf{x}')\, \sigma_L^2 \!\left(r_L(\textbf{x},\textbf{x}') - \textbf{r}_L(\textbf{x})^\top \!\left(R_L + \eta_L I_{N_L}\right)^{-1}\textbf{r}_L(\textbf{x}')\right) + \sigma_H^2\, r_H(\textbf{x},\textbf{x}') \\
&\quad - \textbf{k}_{AR}(\textbf{x})^\top \!\left(K_{AR} + \sigma^2_{\varepsilon,H} I_{N_H}\right)^{-1}\textbf{k}_{AR}(\textbf{x}') \\
&= \rho(\textbf{x})\rho(\textbf{x}')\, v_{Y_L}(\textbf{x},\textbf{x}') + \sigma_H^2\, r_H(\textbf{x},\textbf{x}') - \textbf{k}_{AR}(\textbf{x})^\top \!\left(K_{AR} + \sigma^2_{\varepsilon,H} I_{N_H}\right)^{-1}\textbf{k}_{AR}(\textbf{x}') \\
&= k_{AR}(\textbf{x},\textbf{x}') - \textbf{k}_{AR}(\textbf{x})^\top \!\left(K_{AR} + \sigma^2_{\varepsilon,H} I_{N_H}\right)^{-1}\textbf{k}_{AR}(\textbf{x}')
= v_{Y_H}(\textbf{x},\textbf{x}').
\end{align*}

We split the residual vector $\textbf{z}_{LH} - F_{LH}\boldsymbol\beta_{LH} = (\textbf{e}_L^\top, \textbf{e}_H^\top)^\top$ with:
\[  
\begin{cases}
    \textbf{e}_L = \textbf{z}^L_{tr} - F_L\boldsymbol\beta_L \\
    \textbf{e}_H = \textbf{z}^H_{tr} - \mathrm{D}_\rho f_L(X_{tr}^H)\boldsymbol\beta_L - F_H\boldsymbol\beta_H.
\end{cases}
\]
Using the expression of $m_{Y_L}$, we can obtain:
\[
\begin{cases}
    \textbf{a}(\textbf{x})^\top A^{-1}\textbf{e}_L
= \rho(\textbf{x})\,\textbf{r}_L(\textbf{x})^\top \!\left(R_L + \eta_L I_{N_L}\right)^{-1}\left(\textbf{z}^L_{tr} - F_L\boldsymbol\beta_L\right)
= \rho(\textbf{x})\!\left(m_{Y_L}(\textbf{x}) - \textbf{f}_L(\textbf{x})^\top \boldsymbol\beta_L\right) \\
B^\top A^{-1}\textbf{e}_L
= \mathrm{D}_\rho\, r_L(X^H_{tr},X^L_{tr}) \left(R_L + \eta_L I_{N_L}\right)^{-1}\left(\textbf{z}^L_{tr} - F_L\boldsymbol\beta_L\right)
= \mathrm{D}_\rho \!\left(\textbf{m}_{Y_L}(X^H_{tr}) - f_L(X_{tr}^H)\boldsymbol\beta_L\right) \\
\textbf{e}_H - B^\top A^{-1}\textbf{e}_L = \textbf{z}^H_{tr} - \mathrm{D}_\rho\,\textbf{m}_{Y_L}(X^H_{tr}) - F_H\boldsymbol\beta_H = \textbf{z}^H_{tr} - \textbf{m}_{AR}.
\end{cases}
\]
Applying \eqref{eq:schur_bilinear} with $\textbf{u} = \textbf{v}_{LH}(\textbf{x})$ and $\textbf{w} = (\textbf{e}_L^\top,\textbf{e}_H^\top)^\top$ yields the equality of the posterior mean functions:
\begin{align*}
m_{Y_H^{KO}}(\textbf{x})
&= \rho(\textbf{x})\,\textbf{f}_L(\textbf{x})^\top\boldsymbol\beta_L + \textbf{f}_H(\textbf{x})^\top\boldsymbol\beta_H
+ \rho(\textbf{x})\!\left(m_{Y_L}(\textbf{x}) - \textbf{f}_L(\textbf{x})^\top \boldsymbol\beta_L\right) \\
&\quad+ \textbf{k}_{AR}(\textbf{x})^\top \!\left(K_{AR} + \sigma^2_{\varepsilon,H} I_{N_H}\right)^{-1}\!\left(\textbf{z}^H_{tr} - \textbf{m}_{AR}\right) \\
&= m_{AR}(\textbf{x}) + \textbf{k}_{AR}(\textbf{x})^\top \!\left(K_{AR} + \sigma^2_{\varepsilon,H} I_{N_H}\right)^{-1}\!\left(\textbf{z}^H_{tr} - \textbf{m}_{AR}\right)
= m_{Y_H}(\textbf{x}),
\end{align*}
which concludes the proof. \hfill $\square$

\end{document}